\documentclass[10pt,letterpaper]{article}
\usepackage[top=0.85in,left=2.75in,footskip=0.75in]{geometry}

\usepackage{float}
\usepackage{graphicx}
\graphicspath{{./}}
\usepackage[export]{adjustbox} 

\usepackage{amsmath,amssymb,verbatim,bm}
\usepackage{booktabs}

\usepackage{changepage}

\usepackage[utf8x]{inputenc}

\usepackage{textcomp,marvosym}

\usepackage{cite}

\usepackage{nameref,hyperref}

\usepackage[right]{lineno}

\usepackage{microtype}
\DisableLigatures[f]{encoding = *, family = * }

\usepackage[table]{xcolor}

\definecolor{violet}{RGB}{0,0,0}

\usepackage{array}

\newcolumntype{+}{!{\vrule width 2pt}}

\newlength\savedwidth

\newcommand\thickhline{\noalign{\global\savedwidth\arrayrulewidth\global\arrayrulewidth 2pt}%
  \hline
  \noalign{\global\arrayrulewidth\savedwidth}}

\raggedright
\usepackage[aboveskip=1pt,labelfont=bf,labelsep=period,justification=raggedright,singlelinecheck=off]{caption}

\makeatletter
\renewcommand{\@biblabel}[1]{\quad#1.}
\makeatother

\usepackage{lastpage,fancyhdr,graphicx}
\usepackage{epstopdf}
\usepackage{xspace}

\fancyheadoffset[L]{2.25in}
\fancyfootoffset[L]{2.25in}
\usepackage{mdframed}
\newmdenv[
  leftmargin = 0pt,
  innerleftmargin = 1em,
  innertopmargin = 0pt,
  innerbottommargin = 0pt,
  innerrightmargin = 0pt,
  rightmargin = 0pt,
  linewidth = 3pt,
  topline = false,
  rightline = false,
  bottomline = false,
  linecolor=violet,
  ]{leftbar}

\newcommand{\bfi}{\bm{\phi}}

\DeclareMathOperator{\rp}{rp}

\DeclareRobustCommand{\dw}[1]{{\Delta w_{#1}}}
\DeclareMathOperator{\drpo}{d_{\rp 1}}

\newtheorem{theorem}{Theorem}

\DeclareRobustCommand{\o}[1]{{\scriptstyle \mathcal{O}} \left(#1\right)}
\DeclareRobustCommand{\O}[1]{{\mathcal{O}} \mspace{-4mu} \left(#1\right)}

\DeclareRobustCommand{\wE}[1]{\widehat{\mathbb{E}}\left[#1\right]}

\newcommand{\bA}{\bm{A}}
\newcommand{\bI}{\bm{I}}
\newcommand{\bM}{\bm{M}}

\newcommand{\bD}{\bm{D}}
\newcommand{\bL}{\bm{L}}
\newcommand{\bcL}{\bm{\mathcal{L}}}

\newcommand{\bP}{\bm{P}}
\newcommand{\bR}{\bm{R}}
\newcommand{\cG}{\mathcal{G}}
\newcommand{\cD}{\mathcal{D}}
\newcommand{\hD}{\widehat{D}}
\newcommand{\hcD}{\widehat{\cD}}

\newcommand{\lamA}{\lambda^{\bA}}
\newcommand{\lamL}{\lambda^{\bL}}
\newcommand{\lamNL}{\lambda^{\bcL}}
\newcommand{\Prob}[1]{\mathbb{P}\left(#1\right)}
\newcommand{\DC}{\textsc{DeltaCon}\xspace}
\newcommand{\NS}{\textsc{NetSimile} }

\newcommand{\eqdef}{\stackrel{\text{\tiny def}}{=}}

\newcommand{\hspc[1]}{\hspace{#1pt}}

\newcommand{\ER}{Erd\H{o}s-R\'enyi\xspace}

\begin{document}
\vspace*{0.2in}

\begin{flushleft}
  {\Large
    \textbf\newline{Metrics for Graph Comparison: A Practitioner's Guide} 
  }
  \newline
  \\
  Peter Wills\textsuperscript{1},
  Fran\c{c}ois G. Meyer\textsuperscript{1*},
  \\
  \bigskip
  \textbf{1} Applied Mathematics, University of Colorado at Boulder, Boulder CO 80305
  \\
  \bigskip
  * fmeyer@colorado.edu

\end{flushleft}
\section*{Abstract}
Comparison of graph structure is a ubiquitous task in data analysis and machine learning, with diverse applications in
fields such as neuroscience \cite{bassett2017,devicofallani14,fornito16}, cyber security
\cite{chen2012,Pasqualetti2013}, social network analysis \cite{Myers2014}, and bioinformatics \cite{Garroway2008}, among
others. Discovery and comparison of structures such as modular communities, rich clubs, hubs, and trees yield insight
into the generative mechanisms and functional properties of the graph.

Often, two graphs are compared via a pairwise distance measure, with a small distance indicating structural similarity
and vice versa. Common choices include spectral distances and distances based on node affinities. However, there has of
yet been no comparative study of the efficacy of these distance measures in discerning between common graph topologies
at different structural scales.

In this work, we compare commonly used graph metrics and distance measures, and demonstrate their ability to discern
between common topological features found in both random graph models and real world networks. We put forward a
multi-scale picture of graph structure wherein we study the effect of global and local structures on changes in distance
measures. We make recommendations on the applicability of different distance measures to the analysis of empirical graph
data based on this multi-scale view. Finally, we introduce the Python library \texttt{NetComp} \cite{netcomp19} that
implements the graph distances used in this work.


\section*{Introduction}
\setcounter{section}{1}\setcounter{subsection}{0} \setcounter{subsubsection}{0} In the era of big
data, comparison and matching are ubiquitous tasks. A graph is a particular type of data structure
that records the interactions between some collection of agents.\footnote{These objects are
  sometimes referred to as ``complex networks;'' we use the mathematician's term ``graph''
  throughout the paper.} This type of data structure relates connections between objects, rather
than directly relating the properties of those objects. The interconnectedness of the object in
graph data disallows many common statistical techniques used to analyze tabular datasets. The need
for new analytical techniques for visualizing, comparing, and understanding graph data has given
rise to a rich field of study \cite{Cook2006,haslbeck2018,liu2019}.

In this work, we focus on tools for pairwise comparison of graphs. Examples of applications include
the two-sample test problem and the change point detection problem.  In the former, we compare two
populations of graphs using a distance statistic, and we experimentally test whether both
populations could be generated by the same probability distribution. In the latter, we consider a
dynamic network formed by a time series of graphs, and the goal is to detect significant changes
between adjacent time steps using a distance \cite{Holme2012}. Both problems require the ability to
effectively compare two graphs. However, the utility of any given comparison method varies with the
type of information the user is looking for; one may care primarily about large scale graph features
such as community structure or the existence of highly connected ``hubs''; or, one may be focused on
smaller scale structure such as local connectivity (i.e. the degree of a vertex) or the ubiquity of
substructures such as triangles.

Existing surveys of graph distances are limited to observational datasets (e.g., \cite{Donnat2018}
and references therein). While authors try to choose datasets that are exemplars of certain classes
of networks (e.g., social, biological, or computer networks), it is difficult to generalize these
studies to other datasets.

In this paper, we take a different approach. We consider existing ensembles of random graphs as
prototypical examples of certain graph \emph{structures}, which are the building blocks of existing
real world networks. We propose therefore to study the ability of various distances to compare two
samples randomly drawn from distinct ensembles of graphs. Our investigation is concerned with the
relationship between the families of graph ensembles, the structural features characteristic of
these ensembles, and the sensitivity of the distances to these characteristic structural features.

The myriad of proposed techniques for graph comparison \cite{Akoglu2015} are severely reduced in
number when one requires the practical restriction that the algorithm run in a reasonable amount of
time on large graphs. Graph data frequently consists of $10^4$ to $10^8$ vertices, and so algorithms
whose complexity scales quadratically with the size of the graph quickly become unfeasible. In this
work, we restrict our attention to approaches where the calculation time scales linearly or
near-linearly with the number of vertices in the graph for sparse graphs.\footnote{Sparsity is,
  roughly, the requirement that the number of edges in a graph of size $n$ be much lower than the
  maximum possible number $n^2/2$; a technical definition is provided below.}

In the past 40 years, many random graph models have been developed that emulate certain features
found in real-world graphs \cite{Barabasi1999, Watts1998}. A rigorous probabilistic study of the
application of graph distances to these random models is difficult because the models are often
defined in terms of a generative process rather than a distribution over the space of possible
graphs. As such, researchers often restrict their attention to very small, deterministic graphs (see
e.g., \cite{Monnig2016}) or to very simple random models, such as that proposed by Erd\H{o}s and
R\'enyi \cite{Erdos1959}. Even in these simple cases, rigorous probabilistic analysis can be
prohibitively difficult. We propose instead a numerical approach where we sample from random graph
distributions and observe the empirical performance of various distance measures.

Throughout the work, we examine the observed results through a lens of global versus local graph
structure. Examples of global structure include community structure and the existence of
well-connected vertices (often referred to as ``hubs''). Examples of local structure include the
median degree in the graph, or the density of substructures such as triangles. Our results
demonstrate that some distances are particularly tuned towards observing global structure, while
others naturally observe multiple scales. In both empirical and numerical experiments, we use
this multi-scale interpretation to understand why the distances perform the way they do on a given
model, or on given empirical graph data.

The paper is structured as follows: in Section \ref{sec:graph-dist-meas}, we introduce the distances
used, and establish the state of knowledge regarding each. In Section \ref{sec:random-graph-models},
we describe the random graph ensembles that are used to evaluate the various distances. We discuss
their structural features, and their respective values as prototypical models for real networks. In
Section \ref{sec:real_networks} we describe three real networks that we use to further study the
performance of the distances. The reader who is already familiar with the graph models and distances
discussed can skip to Section \ref{sec:how_to_measure} to find the description of the contrast
statistic that we use to compare graph populations. Experimental results are briefly described in
sections \ref{sec:results_graph_models} and \ref{sec:results_real_networks}. A detailed discussion
is provided in Sections \ref{discussion_ensembles} and \ref{discussion_real_networks}. Finally,
Section {\bfseries Conclusion} summarizes the work and our recommendations. In the appendix, we
introduce and discuss \texttt{NetComp}, the Python package that implements the distances used to
compare the graphs throughout the paper.
\section*{Methods}
\setcounter{section}{2}\setcounter{subsection}{0} \setcounter{subsubsection}{0}
\subsection{Notation}
We must first introduce the notation used throughout the paper. It is standard
wherever possible.

We denote by $G = (V,E,W)$ a graph with vertex set $V=\{1,\ldots,n\}$ and edge set
$E\subseteq V\times V$. The function $W:E\rightarrow \bR^+$ assigns each edge
$(i,j)$ in $E$ a positive weight that we denote $w_{i,j}$. We call $n=|V|$ the
\textbf{size} of the graph, and denote by $m\eqdef |E|$ the number of edges. For
$i\in V$ and $j\in V$, we say $i\sim j$ if $(i,j)\in E$. The matrix $\bA$ is
called the \textbf{adjacency matrix}, and is defined as
\begin{equation*}
  A_{i,j} \eqdef \left\{ 
    \begin{array}{cl}
      w_{i,j} & \text{if} \,\, i\sim j, \\
      0 & \text{otherwise.}
    \end{array}
  \right.
\end{equation*}

The \textbf{degree} $d_i$ of a vertex is defined as $d_i\eqdef\sum_{j\sim
  i}w_{i,j}$. The \textbf{degree matrix} $\bD$ is the diagonal matrix of degrees,
so $D_{i,i}=d_i$ and $D_{i,j}=0$ for $i\neq j$. The combinatorial \textbf{Laplacian matrix}
(or just Laplacian) of $G$ is given by $\bL \eqdef \bD - \bA$. The \textbf{normalized
  Laplacian} is defined as $\bcL \eqdef \bD^{-1/2} \bL \bD^{-1/2}$, where the
diagonal matrix $\bD^{-1/2}$ is given by
\begin{equation*}
  D^{-1/2}_{i,i} \eqdef \left\{
    \begin{array}{cl}
      1/\sqrt{d_{i}} & \text{if} \,\, d_i\neq 0, \\
      0 & \text{otherwise.}
    \end{array}
  \right.
\end{equation*}

We refer to $\bA$, $\bL$, and $\bcL$ as \textbf{matrix representations} of $G$. These are not the
only useful matrix representations of a graph, although they are some of the most common. For a more
diverse catalog of representations, see \cite{Wilson2008}. 

The \textbf{spectrum} of a matrix is the sorted sequence of eigenvalues. Whether the sequence is
ascending or descending depends on the matrix in question. We denote the $k^\text{th}$ eigenvalue of
the adjacency matrix by $\lamA_k$, where we order the eigenvalues in descending order
\begin{equation}
  \lamA_1 \geq \lamA_2\geq\ldots\geq \lamA_n.
\end{equation}
We denote by $\lamL_k$ the $k^\text{th}$ eigenvalue of the Laplacian matrix, and we order these
eigenvalues in \textbf{ascending} order, so that
\begin{equation}
  0=\lamL_1\leq \lamL_2\leq\ldots\leq\lamL_n.
\end{equation}
We similarly denote the $k^\text{th}$ eigenvalue of the normalized Laplacian by $\lamNL_k$, with
\begin{equation}
0=  \lamNL_1\leq \lamNL_2 \leq \ldots\leq \lamNL_n,
\end{equation}
and we denote by $\bfi_k$ the corresponding eigenvector. 

The significance of this convention is that the index $k$ of an eigenvalue $\lambda_k$ always
encodes the frequency of the corresponding eigenvector. To wit, the eigenvector associated with
either $\lamA_k$ or $\lamL_k$ experiences about $k$ oscillations on the graphs, with $k+1$ nodal
domains \cite{biyikoglu2007}.\\ 
~\\

Two graphs $G$ and $G'$ are \textbf{isomorphic} if and only if there exists a map between their
vertex sets under which the two edge sets are equal; we write $G \cong G'$. If we denote by $\bA$
and $\bA'$ the adjacency matrices of $G$ and $G'$ respectively, then $G \cong G'$ if and only if
there exists a permutation matrix $\bP$ such that $\bA' = \bP^T \bA \bP$.

We say that a distance $d$ requires \textbf{node correspondence} when there exist graphs $G$, $G'$,
and $H$ such that $G\cong G'$ but $d(G,H) \neq d(G',H)$. Intuitively, a distance requires node
correspondence when one must know some meaningful mapping between the vertex sets of the graphs
under comparison.
\subsection{Graph distance measures
\label{sec:graph-dist-meas}}
Let us begin by introducing the distances that we study in this paper, and discussing the state of the
knowledge for each. We have chosen both standard and cutting-edge distances, with the requirement
that the algorithms be computable in a reasonable amount of time on large, sparse graphs. In
practice, this means that the distances must scale linearly or near-linearly in the size in the
graph.

We refer to these tools as ``distance measures,'' as many of them do not satisfy the technical
requirements of a metric. Although all are symmetric, they may fail one or more of the other
requirements of a mathematical metric. This can be very problematic if one hopes to perform rigorous
analysis on these distances, but in practice it is not significant. Consider the requirement of
identity of indiscernible, in which $d(G,G')=0$ if and only if $G=G'$. We rarely encounter two
graphs where $d(G,G')=0$; we are more frequently concerned with an approximate form of this
statement, in which we wish to deduce that $G$ is similar to $G'$ from the fact that $d(G,G')$ is
small.

The distance measures we study divide naturally into two categories, that we now describe. These
categories are not exhaustive; many distance measures (including one we employ in the experiments)
do not fit neatly into either category. Akoglu et al. \cite{Akoglu2015} provide an alternative
taxonomy; our taxonomy refines a particular group of methods they refer to as
``feature-based''.\footnote{Note that the authors in \cite{Akoglu2015} are classifying anomaly
  detection methods in particular, rather than graph comparison methods in general.}
\subsubsection{Spectral distances
\label{sec:spectral-distances}}
Let us first discuss spectral distances. We briefly review the necessary background; see
\cite{Wilson2008} for a good introduction to spectral methods used in graph comparison.

We first define the adjacency spectral distance; the Laplacian and normalized Laplacian spectral
distances are defined similarly. Let $G$ and $G'$ be graphs of size $n$, with adjacency spectra
$\lamA$ and $\lambda^{\bA'}$, respectively. The \textbf{adjacency spectral distance} between the two
graphs is defined as
\begin{equation*}
  d_{\bA}(G,G') \eqdef \sqrt{ \sum_{i=1}^n \left(\lamA_i - \lambda^{\bA'}_i\right)^2 },
\end{equation*}
which is just the distance between the two spectra in the $\ell_2$ metric. We could use any $\ell_p$
metric here, for $p\in[0,\infty]$. The choice of $p$ is informed by how much one wishes to emphasize
outliers; in the limiting case of $p=0$, the metric returns the measure of the set over that the two
vectors are different, and when $p=\infty$ only the largest element-wise difference between the two
vectors is returned. Note that for $p<1$ the $\ell_p$ distances are not true metrics (in particular,
they fail the triangle inequality) but they still may provide valuable information. For a more
detailed discussion on $\ell_p$ norms, see \cite{Rudin1991}.

The Laplacian and normalized Laplacian spectral distances $d_{\bL}$ and $d_{\bcL}$ are defined in
the exact same way. In general, one can define a spectral distance for any matrix representation of
a graph; for results on more than just the three we analyze here, see \cite{Wilson2008}. We note
that spectral distances do not require node correspondence.

An important property of the normalized Laplacian spectral distance is that it can be used to
compare graphs of different sizes (see e.g., \cite{delang14}).

In practice, it is often the case that only the first $k$ eigenvalues are compared, where $k\ll
n$. We still refer to such truncated spectral distances as spectral distances. When using spectral
distances, it is important to keep in mind that the adjacency spectral distance compares the
\textit{largest} $k$ eigenvalues, whereas the Laplacian spectral distances compare the
\textit{smallest} $k$ eigenvalues. Comparison using the first $k$ eigenvalues $\lamA_k$ for small
$k$ allows one to focus on the community structure of the graph, while ignoring the local structure
of the graph \cite{Lee2014}. Inclusion of the highest-$k$ eigenvalues $\lamA_k$ allows one to discern
local features as well as global. This flexibility allows the user to target the particular scale at
which she wishes to study the graph, and is a significant advantage of the spectral distances.

The three spectral distances used here are not true metrics. This is because there exist graphs $G$
and $G'$ that are co-spectral but not isomorphic. That is to say, adjacency cospectrality occurs
when $\lamA_i = \lambda^{\bA'}_i$ for all $i=1,\ldots,n$, so $d_{\bA}(G,G')=0$, but $G\ncong
G'$. Similar notions of cospectrality exist for all matrix representations; graphs that are
co-spectral with respect to one matrix representation are not necessarily co-spectral with respect
to other representations.

Little is known about cospectrality, save for some computational results on small graphs
\cite{Haemers2004} and trees \cite{Wilson2008}. Schwenk proved that a sufficiently large tree nearly
always has a co-spectral counterpart \cite{Schwenk1973}. This result was extended recently to
include a wide variety of random trees \cite{Bhamidi2012}. However, results such as these are not of
great import to us; the graphs examined are large enough that we do not encounter cospectrality in
our numerical experiments. A more troubling failure mode of the spectral distances would be when the
distance between two graphs is very small, but the two graphs have important topological
distinctions. In Section {\bfseries Discussion}, we provide further insight into the effect of
topological changes on the spectra of some of the random graph models we study.

The consideration above addresses the question of how local changes affect the overall spectral
properties of a graph. Some limited computational studies have been done in this direction. For
example, Farkas et al. \cite{Farkas2001} study the transition of the adjacency spectrum of a small
world graph as the disorder parameter increases. As one might expect, the authors in
\cite{Farkas2001} observe the spectral density transition from a highly discontinuous density (which
occurs when the disorder is zero and the graph is a ring-like lattice) to Wigner's famous
semi-circular shape \cite{Wigner1958} (which occurs when the disorder is maximized, so that the
graph is roughly equivalent to an uncorrelated random graph.)

From an analytical standpoint, certain results in random matrix theory inform our understanding of
fluctuations of eigenvalues of the uncorrelated random graph (see Section {\bfseries Random Graph
  Models} for a definition). These results hold asymptotically as we consider the $k^\text{th}$
eigenvalue of a graph of size $n$, where $k = \alpha n$ for $\alpha\in(0,1]$. In this case, O'Rourke
\cite{ORourke2010} has shown that the the eigenvalue $\lambda_k$ is asymptotically normal with
asymptotic variance $\sigma^2(\lambda_k) = C(\alpha) \log n/n$. An expression for the constant
$C(\alpha)$ is provided; see Remark 8 in \cite{ORourke2010} for the detailed statement of the
theorem. This result can provide a heuristic for spectral fluctuations in some random graphs, but
when the structure of these graphs diverges significantly from that of the uncorrelated random
graph, then results such as these become less informative.


Another common question is that of interpretation of the spectrum of a given matrix representation
of a graph.\footnote{``Spectral structure'' might refer to the overall shape of the spectral
  density, or the value of individual eigenvalues separated from the bulk.} How are we to understand
the shape of the empirical distribution of eigenvalues? Can we interpret the eigenvalues which
separate from this bulk in a meaningful way? The answer to this question depends, of course, on the
matrix representation in question. Let us focus first on the Laplacian matrix $\bL$, the
interpretation of that is the clearest.

The first eigenvalue of $\bL$ is always $\lamL_1=0$, with the eigenvector being the vector of all
ones, $\mathbf{1}\in\mathbb{R}^n$. It is a well-known result that the multiplicity of the zero
eigenvalue is the number of connected components of the graph, i.e. if $0=\lamL_k<\lamL_{k+1}$, then
there are precisely $k$ connected components of the graph \cite{Chung1997}. Furthermore, in such a
case, the first $k$ eigenvectors can be chosen to be the indicator functions of the components.
There exists a relaxed version of this result: if the first $k$ eigenvalues are very small (in a
sense properly defined), then the graph can be strongly partitioned into $k$ clusters (see
\cite{Lee2014} for the rigorous formulation of the result). This result justifies the use of the
Laplacian in spectral clustering algorithms, and can help us understand the interplay between the
presence of small eigenvalues and the presence of communities in the ensembles of random graphs
studied in Section \ref{sec:SBM_results}.

The eigenvalues of the Laplacian can be interpreted as vibrational frequencies in a manner similar
to the eigenvalues of the continuous Laplacian operator $\nabla^2$. To understand this analogy,
consider the graph as embedded in a plane, with each vertex representing an oscillator of mass one
and each edge a spring with elasticity one. Then, for small oscillations perpendicular to the plane,
the Laplacian matrix is precisely the coupling matrix for this system, and the eigenvalues give
the square of the normal mode frequencies, $\omega_i = \sqrt{\lamL_i}$. For a more thorough
discussion of this interpretation of the Laplacian, see \cite{Friedman2004}.

Maas \cite{Maas1985} suggests a similar interpretation of the spectrum of the adjacency matrix
$\bA$. Consider the graph as a network of oscillators, embedded in a plane as previously
discussed. Additionally, suppose that each vertex is connected to so many external non-moving points
(by edges with elasticity one) so that the graph becomes regular with degree $d$. The frequencies of
the normal modes of this structure then connect to the eigenvalues of $\bA$ via
$\omega_i=\sqrt{d-\lamA_i}$.\footnote{If the graph is already regular with degree $d$, then this
  interpretation is consistent with the previous, since the eigenvalues of $\bL = d\bI - \bA$ are
  just $\lamL_i = d-\lamA_i$.}

\subsubsection{Matrix distances}
The second class of distances we discuss are called \emph{matrix distances}, and consist of direct
comparison of the structure of pairwise affinities between vertices in a graph (see
\cite{Monnig2016} for a detailed discussion on matrix distances). These affinities are frequently
organized into matrices, and the matrices can then be compared, often via an entry-wise $\ell_p$
norm. Matrix distances all require node correspondence.

We have discussed spectral methods for measuring distances \textit{between} two graphs; to introduce
the matrix distances, we begin by focusing on methods for measuring distances \textit{on a graph};
that is to say, the distance $\delta(v,w)$ between two vertices $v,w\in V$. Just a few examples of
such distances include the shortest path distance \cite{Moore1959}, the effective graph resistance
\cite{Ellens2011}, and variations on random-walk distances \cite{Haveliwala2003}. Of those listed
above, the shortest path distance is the oldest and the most thoroughly studied; in fact, it is so
ubiquitous that ``graph distance'' is frequently used synonymously with shortest path distance
\cite{Goddard2011}.

There are important differences between the distances $\delta$ that we might choose. The shortest
path distance considers only a single path between two vertices. In comparison, the effective graph
resistance takes into account all possible paths between the vertices, and so measures not only the
length, but the \emph{robustness} of the communication between the vertices.

How do these distances \textit{on a graph} help us compute distances \textit{between graphs}? Let us
denote by $\delta:V\times V\rightarrow \mathbb{R}$ a generic distance on a graph. We need assume
very little about this function, besides it being real-valued; in particular, it need not be
symmetric, and we can even allow $\delta(v,v) \neq 0$.\footnote{When we say ``distance'' we
  implicitly assume that smaller values imply greater similarity; however, we can also carry out
  this approach with a similarity score, in which larger values imply greater similarity.} Recalling
that the vertices $v\in V=\{1,\ldots,n\}$ are labelled with natural numbers, we can then construct a
matrix of pairwise distances $\bM$ via $M_{i,j} \eqdef \delta(i,j)$. The idea behind what we refer
to as \textbf{matrix distances} is that this matrix $\bM$ carries important structural information
about the graph.

Consider two graphs $G= (V,E)$ and $G'=(V,E')$ defined on the same vertex set. Given a graph
distance $\delta(\cdot,\cdot)$, let $\bM$ and $\bM'$ be the matrices of pairwise distances between
vertices in the graph $G$ and $G'$ respectively. We define the distance $d$ induced by $\delta$
between $G$ and $G'$ as follows,
\begin{equation}
  \label{eq:1}
  d(G,G') \eqdef \|\bM-\bM'\|,
\end{equation}
where $\|\cdot\|$ is a norm we are free to choose.\footnote{We could use
  metrics, or even similarity functions here, although that may cause the
  function $d$ to lose some desirable properties.}

Let us elucidate a specific example of such a distance; in particular, we show how the edit distance
conforms to this description. Let $\delta(v,w)$ be defined as
\begin{equation}
  \label{eq:2}
  \delta(v,w) = \left\{
    \begin{array}{cl}
      1 & \text{if} \,\, v \sim w, \\
      0 & \text{else.}
    \end{array}
  \right.
\end{equation}
Then the matrix $\bM$ is just the adjacency matrix $\bA$. If we use the norm
\begin{equation}
  \label{eq:3}
  \|\bM\| = \sum_{i,j=1}^n |M_{i,j}|,
\end{equation}
then we call the resulting distance $d(G,G') \eqdef \|\bA-\bA'\|$ the \textbf{edit distance}.

Of course, the usefulness of such a distance is directly dependent on how well
the matrix $\bM$ reflects the topological structure of the graph. The edit
distance focuses by definition on local structure; it can only see changes at
the level of edge perturbations. If significant volume changes are happening in
the graph, then the edit distance detects these changes, as do other matrix
distances.

To compensate for such trivial first order changes (changes in volume) we match the expected volume
of the models under comparison (see Section~\ref{sec:results_graph_models}). We can then study
whether distances can detect structural changes.

We also implement the resistance-perturbation distance, first discussed in
\cite{Monnig2016}. This distance takes the effective graph resistance $R(u,v)$,
defined in \cite{Ellens2011}, as the measure of vertex affinity. This results in
a (symmetric) matrix of pairwise resistances $\bR$. The resistance-perturbation
distance (or just resistance distance) is based on comparing these two matrices
in the entry-wise $\ell_1$ norm given in Equation \eqref{eq:3}. 

Unlike the edit distance, the resistance distance is designed to detect changes in connectivity
between graphs. A recent work \cite{Wills2017} discusses the efficacy of the resistance distance in
detecting community changes.

Finally, we study \textsc{DeltaeCon}, a distance based on the fast belief propagation method of
measuring node affinities \cite{Koutra2014}. To compare graphs, this method uses the fast
belief propagation matrix
\begin{equation}
  \label{eq:5}
  \mathbf{S} \eqdef [\bI + \epsilon^2\bD - \epsilon \bA]^{-1},
\end{equation}
and compares the two representations $\mathbf{S}$ and $\mathbf{S}'$ via the
Matusita difference,
\begin{equation}
  \label{eq:6}
  d(G,G') = \sqrt{\sum_{i,j}\left(\sqrt{S_{i,j}} - \sqrt{S'_{i,j}}\right)^2}.
\end{equation}
Note that the matrix $\mathbf{S}$ can be rewritten in a matrix power series as
\begin{equation}
  \label{eq:7}
  \mathbf{S}\approx \bI + \epsilon \bA + \epsilon^2 (\bA^2-\bD) + \ldots
\end{equation}
and so takes into account the influence of neighboring vertices in a weighted manner, where
neighbors separated by paths of length $k$ have weight $\epsilon^k$. Fast belief propagation is
designed to model the diffusion of information throughout a graph \cite{Koutra2011}, and so should
in theory be able to perceive both global and local structures. Although empirical tests are
performed in \cite{Koutra2014}, no direct comparison to other modern methods is presented.
\subsubsection{Feature-based distances}
These two categories do not cover all possible methods of graph comparison. The computer science
literature explores various other methods (e.g., see \cite{Akoglu2015}, Section 3.2 for a comprehensive
review), and other disciplines that apply graph-based techniques often have their own idiosyncratic
methods for comparing graphs extracted from data.

One possible method for comparing graphs is to look at specific ``features'' of the graph, such as
the degree distribution, betweenness centrality distribution, diameter, number of triangles, number
of $k$-cliques, etc. For graph features that are vector-valued (such as degree distribution) one
might also consider the vector as an empirical distribution and take as graph features the sample
moments (or quantiles, or statistical properties). A \emph{feature-based distance} is a distance
that uses comparison of such features to compare graphs.

Of course, in a general sense, all methods discussed so far are feature based; however, in the
special case where the features occur as values over the space $V\times V$ of possible node
pairings, we choose to refer to them more specifically as \emph{matrix distances}. Similarly, if the
feature in question is the spectrum of a particular matrix realization of the graph, we call the
method a spectral distance.

In \cite{Berlingerio2012}, a feature-based distance called \textsc{NetSimile} is proposed, which
focuses on local and egonet-based features (e.g., degree, volume of egonet as fraction of maximum
possible volume, etc.). If we are using $k$ features, the method aggregates a feature-vertex matrix
of size $k\times n$. This feature matrix is then reduced to a ``signature vector'' (a process the
authors in \cite{Berlingerio2012} call ``aggregation'') that consists of the mean, median,
standard deviation, skewness, and kurtosis of each feature. These signature vectors are then
compared in order to obtain a measure of distance between graphs.

In the neuroscience literature in particular, feature-based methods for comparing graphs are popular
\cite{bassett2008,kaiser2011}. In \cite{Heuvel2013}, the authors use graph features such as modularity, shortest
path distance, clustering coefficient, and global efficiency to compare functional connectivity
networks of patients with and without schizophrenia. Statistics of these features for the control
and experiment groups are aggregated and compared using standard statistical techniques.

We implement \textsc{NetSimile} as a prototypical feature-based method. It is worth noting that the
general approach could be extended in almost any direction; any number of features could be used
(which could take on scalar, vector, or matrix values) and the aggregation step can include or omit
any number of summary statistics on the features, or can be omitted entirely. We implement the
method as it is originally proposed, with the caveat that calculation of many of these features is
not appropriate for large graphs, as they cannot be computed in linear or near-linear time. A
scalable modification of \textsc{NetSimile} would utilize features that can be calculated (at least
approximately) in linear or near-linear time.
  \subsubsection{Learning Graph Kernels \label{DGK}}
  Given the diversity of structural features in graphs, and the difficulty of designing by hand the
  set of features that optimizes the graph embedding, several researchers have proposed recently to
  learn the embedding from massive datasets of existing networks. Such algorithms learn an embedding
  \cite{goyal18} from a set of graphs into Euclidean space, and then compute a notion of similarity
  between the embedded graphs (e.g., \cite{scarselli08,yanardag15,li19} and references therein). The
  metric that is learnt can be tailored to a specific application (e.g.,
  \cite{coley17,fout17,gilmer17,ktena17,li19,preuer18,yang19}).

All these approaches rely on the extension of convolutional neural networks to non Euclidean
structures, such as manifolds and graphs (e.g., \cite{kipf17,li17,monti16,scarselli09} and
references therein). The core scientific question becomes: how does one implement the convolution
units that are in the network?  Two methods have been proposed. The first method performs the
convolution in the spectral domain \cite{bruna13},(defined by the eigenspace of the graph
Laplacian). These data-dependent convolutions can be performed directly in the spatial domain (using
polynomials of the Laplacian \cite{defferrard16}) or in the spectral domain (in the eigenspace of
the Laplacian). Purely ``in-the-graph'' methods have also been proposed where the convolution is
implemented using an aggregation process (e.g., \cite{gilmer17,battaglia18,zhou18} and references
therein).

Graph kernels \cite{kriege19} are typically not injective (two graphs can be perfectly similar
without being the same), and rarely satisfies the triangular inequality. There have been some recent
attempts at identifying the classes of kernels that are injective \cite{xu18,morris19}. The question
can be rephrased in terms of how expressive is the embedding from the space of graphs to Euclidean
space, i.e. how often do two distinct graphs are mapped to same point \cite{oneto17}. The authors in
\cite{xu18,morris19} have proved that graph neural networks are as expressive as the
Weisfeiler-Lehman graph isomorphism test: if two graphs are mapped to distinct points by the
embedding, then the Weisfeiler-Lehman graph test would consider these graphs to be distinct (non
isomorphic).
  \color{violet}
  \subsubsection{
    Comparing Graphs of Different Sizes}
  \color{violet} The distance measures described in the previous paragraphs are defined for two graphs that have the
  same size. In practice, one often needs to compare graphs of different sizes. Inspired by the rich connections
  between graph theory and geometry, one can define a notion of distance between any two graphs by extending the notion
  of distance between metric spaces \cite{Gromov07}. The construction proceeds as follows: each graph is represented as
  a metric space, wherein the metric is simply the shortest distance on the graph. Two graphs are equivalent if there
  exists an isomorphism between the graph -- represented as metric spaces. Finally, one can define a distance between
  two graphs $G_1$ and $G_2$ (or rather between the two classes of graph isometric to $G_1$ and $G_2$ respectively) by
  considering standard notions of distances between isometry classes of metric spaces \cite{Berger03}. Examples of such
  distances include the Gromov-Hausdorff distance \cite{Berger03}, the Kantorovich-Rubinstein distance and the Wasserstein
  distance \cite{Villani08}, which both require that the metric spaces be equiped with probability measures.
    The Gromov-Hausdorff distance computes the infimum of the Hausdorff distance between the  isometric embeddings of
    two metric spaces into a common one. In plain English, this distance measures the residual error after trying to
    ``optimally align'' two metric spaces using deformations of these spaces that preserve distances
    (isometries). Because the search for the optimal alignment (embedding) is over such a vast space of functions, the
    Gromov-Hausdorff does not lend itself to practical applications (but see \cite{memoli05}).

    On the other hand, the Wasserstein-Kantorovich-Rubinstein distance, also known as the ``Earth Mover's distance'' in the
    engineering literature, has been used extensively in probability and pattern recognition (e.g.,
    \cite{memoli11,panaretos19,peyre19} and references therein). The Wasserstein distance can be interpreted as the cost
    of transporting a measure from one metric space to a second measure defined on a second metric space; the cost
    increases with the distance between the metric spaces and the proportion of the measure that needs to be
    transported. These concepts have just recently been applied to the case of measuring distances between graphs. Given
    a graph $G$, one can associate a measure on the graph (e.g., defined by a histogram of the degrees
    \cite{wu19a,xu19}, a Gaussian measure with a covariance matrix given by the pseudo-inverse of the graph Laplacian
    \cite{maretic19}, or a uniform measure on the graph \cite{titouan19}), and a notion of cost between nodes (e.g., the
    Bures distance \cite{maretic19}, the shortest distance between two nodes \cite{titouan19} assuming the node
    correspondence between the graphs has been established).

    The computational complexity of the estimation of the Wasserstein distance remains prohibitively high for large
    graph: the cost is $m n^2 + m^2n$, where $m$ is the number of edges, and $n$ is the number of nodes. A closed form
    expression of the Wasserstein distance can be derived when the measure on each graph is a Gaussian measure
    \cite{maretic19}. In this case the Wasserstein distance is the Bures distance between their respective covariance
    matrices. This computation is further simplified when the covariance matrices are diagonal, since the Bures distance
    becomes then the Hellinger distance (e.g., \cite{bhatia19a,bhatia19c} and references therein). The rich connection
    between distances between metric spaces, optimal transport, and metrics on the cone of positive semidefinite
    matrices is clearly beyond the scope of the current study; it will certainly provide interesting avenues for future
    studies.

    The relevance of the current study to this burgeoning research area stems from the exploration of the relationship
    between the structural features characteristic of several graph ensembles and the sensitivity of the distances to
    these features. The distributions associated with these features can then be used to define a 
    probability measure associated with a given graph (e.g., \cite{rezaeinia19} where the distribution of hitting times
    is used to characterize a functional brain connectivity network).

\subsection{Computational Efficiency}
\subsubsection{Algorithmic Complexity}

In many interesting graph analysis scenarios, the sizes of the graphs to be analyzed are on the
order of millions or even billions of vertices. For example, the social network defined by Facebook
users has over 2.3 billion vertices as of 2018. In scenarios such as these, any algorithm of
complexity $\O{n^2}$ becomes unfeasible; although in principle it is possible that the constant
hidden in $\O{}$ would be so small it would make up for the $n^2$ term in the complexity, in
practice this is not the case. This motivates the requirement that algorithms be of near-linear
complexity. When the complexity of the distance depends on the graph volume $m$, we assume that the
graph is sparse and $m$ is a linear function (up to a logarithmic factor) of the size $n$.

This challenge motivates the previously stated requirement that all algorithms be of linear or
near-linear complexity. We say an algorithm is \textbf{linear} if it is $\O{n}$; it is
\textbf{near-linear} if it is $\O{n \log a_n }$ where $a_n$ is asymptotically bounded by a
polynomial. We use the notation $a_n = \O{b_n}$ in the standard way; for a more thorough discussion
of algorithmic complexity, including definitions of the Landau notations, see
\cite{Papadimitriou2003}.

Table~\ref{tab:complexity} displays the algorithmic complexity of each distance measure we
compare. We assume that factors such as graph weights and quality of approximation are held
constant, leading to simpler expressions here than appear in cited references. Spectral distances
have equivalent complexity, since they all all amount to performing an eigendecomposition on a
symmetric real matrix. For \DC and the resistance distance, there are approximate algorithms as well
as exact algorithms; we list the complexity of both. Although we use the exact versions in the 
experiments, in practice the approximate version would likely be used if the graphs to be compared
are large.

Of particular interest are the highly parallelizable randomized algorithms which can allow for
extremely efficient matrix decomposition. In \cite{Halko2011}, the authors review many such
algorithms, and discuss in particular their applicability to determining principal eigenvalues. The
computation complexity in Table \ref{tab:complexity} for the spectral distances is based on their
simplified analysis of the Krylov subspace methods, that states that the approach is
$\O{kT_\text{mult} + (m+n)k^2}$, where $T_\text{mult}$ is the cost of matrix-vector multiplication
for the input matrix. Since the input matrices are sparse, $T_\text{mult} = \O{n}$, and
$m+n = \O{n}$. Although the eigensolver uses the implicitly restarted Arnoldi method, if
implementing such a decomposition on large matrices, the use of a randomized algorithm could lead to
a significant increase in efficiency.

\begin{table}[H]
  \centering
  \begin{tabular}{|l|l|c|}
    \hline
    \textbf{Distance Measure} & \textbf{Complexity} & \textbf{Ref.} \\
    \thickhline
    Edit Distance & $\O{m}$ & \footnote{The edit distance, as we define it,
                              consists of subtracting sparse matrices, and thus
                              an efficient implementation scales with the
                              number of entries in the matrices in question.}\\
    Resistance Distance (Exact) & $\O{n^2}$ & \cite{Monnig2016} \\
    Resistance Distance (Approximate) & $\O{m}$ &
                                                  \cite{Monnig2016} \\
    DeltaCon (Exact) & $\O{n^2}$ & \cite{Koutra2014} \\
    DeltaCon (Approximate) & $\O{m}$ & \cite{Koutra2014} \\
    NetSimile & $\O{n \log n}$ & \cite{Berlingerio2012} \\
    Spectral Distance& $\O{nk^2}$ &  \cite{Halko2011}\\
    \hline
  \end{tabular}
  \caption{Distance measures and complexity. The size (of the larger)  graph is $n$; the number of
    edges is $m$. For the spectral decomposition, $k$ denotes the number of principal
    eigenvalues we wish to find. 
  \label{tab:complexity}}
\end{table}

\subsubsection{
  Comparison of Runtimes on Graphs on Small Graphs}

  \color{violet}
In Section \ref{sec:results_graph_models}, we perform the experiments on small graphs,
consisting of only 1,000 nodes.

In application, the graphs under comparison can vary from hundreds up to billions of nodes. We focus on smaller graphs
primarily so that the computation of the distances is tractable even on a small personal computer.

Of course, the time it takes to calculate a given distance depends highly on the implementation of that distance. The
runtimes reported below use the implementations in NetComp \cite{netcomp19}. These implementations are not highly
optimized; spectral calculations depends on the standard spectral solvers that come with \texttt{scipy}, a standard
computational package in Python. These leverage sparse data structures when available.

For the resistance distance and DeltaCon, the distance has both an exact form which has $\O{n^2}$ complexity, and an
approximate form which has $\O{m}$ complexity. We use the exact forms in our calculations, and these are the forms
implemented in NetComp \cite{netcomp19}. For DeltaCon, the approximate form is implemented in MATLAB, and the code is
available on the author's website, \url{http://web.eecs.umich.edu/~dkoutra/}. For the resistance distance, the authors
of \cite{Monnig2016} have released an implementation of the approximate resistance distance in MATLAB, which can be
found on GitHub at \url{https://github.com/natemonnig/Resistance-Perturbation-Distance}. We hope to include Python
implementations of these fast approximate distances in NetComp in the near future.

Table \ref{tab:runtimes} shows the results of our runtime experiments. We compare mean and standard deviations of
runtimes for the various distances. 
\color{violet}
These are computed on small graphs, of size $n=100$, $300$, and $1,000$.
\begin{table}[H]
  \centering
  \begin{tabular}{|l|c|c|}
    \hline
    \textbf{Distance Measure} & \textbf{Computational Time ($n=100$)} \\
    \thickhline
    Edit Distance                   & $8.2 \times 10^{-5} \pm 4.5 \times 10^{-5}$ \\
    DeltaCon                        & $3.1 \times 10^{-3} \pm 7.4 \times 10^{-4}$ \\
    Resistance Dist.                & $7.5 \times 10^{-3} \pm 1.4 \times 10^{-3}$ \\
    Spectral (Adjacency)            & $1.1 \times 10^{-2} \pm 1.1 \times 10^{-3}$ \\
    Spectral (Laplacian)            & $1.2 \times 10^{-2} \pm 4.7 \times 10^{-3}$ \\
    Spectral (Normalized Laplacian) & $1.5 \times 10^{-2} \pm 9.4 \times 10^{-4}$ \\
    NetSimile                       & $2.3 \times 10^{-1} \pm 6.3 \times 10^{-2}$ \\
    \hline
  \end{tabular}
  
  \begin{tabular}{|l|c|c|}
    \hline
    \textbf{Distance Measure} & \textbf{Computational Time ($n=300$)} \\
    \thickhline
    Edit Distance                   & $5.7 \times 10^{-4} \pm 9.9 \times 10^{-4}$ \\
    DeltaCon                        & $1.4 \times 10^{-2} \pm 6.6 \times 10^{-3}$ \\
    Resistance Dist.                & $8.8 \times 10^{-2} \pm 5.4 \times 10^{-2}$ \\
    Spectral (Adjacency)            & $1.5 \times 10^{-1} \pm 8.9 \times 10^{-3}$ \\
    Spectral (Laplacian)            & $1.5 \times 10^{-1} \pm 9.8 \times 10^{-3}$ \\
    Spectral (Normalized Laplacian) & $1.6 \times 10^{-1} \pm 6.6 \times 10^{-3}$ \\
    NetSimile                       & $5.5 \times 10^{-1}\pm 1.1 \times 10^{-2}$ \\
    \hline
  \end{tabular}

  \centerline{\color{violet}
  \begin{tabular}{|l|c|c|}
    \hline
    \textbf{Distance Measure} & \textbf{Computational Time ($n=1,000$)} \\
    \thickhline
    Edit Distance                   & $4.2 \times 10^{-3} \pm 1.5 \times 10^{-3} $ \\
    DeltaCon                        & $8.8 \times 10^{-2} \pm 6.4 \times 10^{-3}$ \\
    Resistance Dist.                & $5.9 \times 10^{-1} \pm 4.3 \times 10^{-2}$ \\
    Spectral (Adjacency)            & $1.3 \pm 5.5 \times 10^{-1}$ \\
    Spectral (Laplacian)            & $1.3 \pm 1.6 \times 10^{-1}$ \\
    Spectral (Normalized Laplacian) & $1.4 \pm 3.7 \times 10^{-1}$ \\
    NetSimile                       & $2.5  \pm 1.8 \times 10^{-1}$ \\
    \hline
  \end{tabular}
}
  \caption{Runtimes for distance various distance measures, for graphs of size $n=100$
    and $n=300$. Each distance is calculated $N=500$ times. Each sample generates two
    \ER random graphs with parameter $p=0.15$, and times the calculation of the distance
    between the two graphs. All distances are implemented in the \texttt{NetComp}
    library, which can be found on GitHub at \cite{netcomp19}.
    \label{tab:runtimes}}
\end{table}

As one might expect, the edit distance is by far the most efficient, as it is simply a difference between and summation
over two sparse matrices. NetSimile is notably slow in our experiments. This is due to inefficient implementation - most
of the work of calculating the various metrics used by NetSimile is done by leveraging NetworkX, a common network
analysis library in Python. Although NetworkX is very simple and clear to work with, it is not designed for maximal
efficiency or scalability, as is evidenced by the above experiments.

We believe it is valuable for the user to get a rough estimate of the efficiency of the easily-available implementations
of the distances discussed in this work. However, much more efficient implementations are possible for each given
distance; these implementations must be carefully designed to be optimal for the particular use-case. A thorough
empirical comparison of the runtimes of optimized implementations of each of these distances would be very illuminating,
but would require considerable care in order to be done equitably, and is well beyond the scope of this work.

\subsection{Random graph models
\label{sec:random-graph-models}}
Random graph models have long been used as a method for understanding topological properties of
graph data that occurs in the world. The uncorrelated random graph model of Erd\H{o}s and R\'enyi
\cite{Erdos1959} is the simplest model, and provides a null model akin to white noise. This
probabilistic model has been analysed thoroughly \cite{Ballobas2001}. Unfortunately, the uniform
topology of the model does not accurately model empirical graph data. The stochastic blockmodel is
an extension of the uncorrelated random graph, but with explicit community structure reflected in
the distribution of edge density.

Models such as preferential attachment \cite{Barabasi1999} and the Watts-Strogatz model
\cite{Watts1998} have been designed to mimic properties of observed graphs. Very little can be said
about these models analytically, and thus much of what is understood about them is
computational. The two-dimensional square lattice is a quintessential example of a highly structured
and regular graph.

Finally, we restrict the present study to unlabelled and undirected graphs, with no
self-loops. Although directed graphs are of great practical importance \cite{Zhou2005}, the
mathematical analysis of directed graphs is far more complex.

Most of the models in this work are sampled via the Python package \textsc{NetworkX}
\cite{Hagberg2008}; details of implementation can be found in the source code of the same. Some of
the models we use are most clearly defined via their associated probability distribution, while
others are best described by a generative mechanism. We introduce the models roughly in order of
complexity.
\subsubsection{The Uncorrelated random graph}
The \textbf{uncorrelated} \ER \textbf{random graph} is a random graph in which each edge exists with
probability $p$, independent of the existence of all others. We denote this distribution of graphs
by $G(n,p)$. The spectral density of the $\lamA$ forms a semi-circular shape, first described by
Wigner \cite{Wigner1958}, of radius $\sqrt{np(1-p)}$, albeit with a single eigenvalue
$\lamA_1\approx np$ separate from the semicircular bulk \cite{Farkas2001}.

We employ the uncorrelated random graph as the null model in many experiments. It is, in some
sense, a ``structureless'' model; more specifically, the statistical properties of each edge and
vertex in the graph are exactly the same. This model fails to produce many of properties observed in
empirical networks, that motivates the use of alternative graph models.
\subsubsection{The Stochastic blockmodel}
One important property of real world networks is community structure. Vertices often form densely
connected communities, with the connection between communities being sparse, or non-existent. This
motivates the use of the \textbf{stochastic blockmodel}. In this model, the vertex set can be
partitioned into two non-overlapping sets $C_1$ and $C_2$ referred to as ``communities'',
\begin{equation}
  V = C_1 \cup C_2.
\end{equation}
Each edge $e = (i,j)$ exists independently with probability $p$ if $i$ and $j$ are in the same
community, and $q$ if $i$ and $j$ are in distinct communities. In this work, we use ``balanced''
communities, whose sizes are equal (up to one vertex in either direction).

The stochastic blockmodel is a prime example of a model that exhibits global structure without any
meaningful local structure. In this case, the global structure is the partitioned nature of the
graph as a whole. On a fine scale, the graph looks like an uncorrelated random graph. We use the
model to determine which distances are most effective at discerning global (and in particular,
community) structure.

The stochastic blockmodel is at the cutting edge of rigorous probabilistic analysis of random
graphs. Abbe et al. \cite{abbe16} have recently proven a strict bound on community
recovery, showing in exactly what regimes of $p$ and $q$ it is possible to detect the communities,
and assign the correct label to each node.

Generalizations of this model exist in which there are $K$ communities of arbitrary
size. Furthermore, each community need not have the same parameter $p$, and each community
\textit{pair} need not have the same parameter $q$. 
\subsubsection{Preferential attachment models}
Another often-studied feature of real world networks is the degree distribution. In practice, the
distribution is estimated using a histogram.

The degree distribution of an uncorrelated random graph is binomial, and so it has tails that decay
exponentially for large graphs (as $n \rightarrow \infty$). However, in real world graphs such as
computer networks, human neural nets, and social networks, the measured degree distribution has a
power-law tail \cite{Barabasi1999}, $\Prob{d}\propto d^{-\gamma}$ where $\gamma \in [2,3]$. Such
distributions are often also referred to as ``scale-free''.

The \textbf{preferential attachment model} is a scale-free random graph model. Although first
described by Yule in 1925 \cite{Yule1925}, the model did not achieve its current popularity until
the work of Barab\'asi and Albert in 1999 \cite{Barabasi1999}.

The model has two parameters, $l$ and $n$. The latter is the size of the graph, and the former
controls the density of the graph. We require that $1\leq l<n$. The generative procedure for
sampling from this distribution proceeds as follows. Begin by initializing a star graph with $l+1$
vertices, with vertex $l+1$ having degree $l$ and all others having degree $1$. Then, for each
$l+1<i\leq n$, add a vertex, and randomly attach it to $l$ vertices already present in the graph,
where the probability of $i$ attaching to $v$ is proportional to to the degree of $v$. We stop once
the graph contains $n$ vertices.

The constructive description of the algorithm does not yield itself to simple analysis, and so less
is known analytically about the preferential attachment model than the uncorrelated random graph or
the stochastic blockmodel (but see \cite{durrett2007,vanderhofstad2016} for some basic properties of
this model). There are few results about the spectrum of the various matrices. In
\cite{Flaxman2003}, the authors prove that if $\lamA_1 \ge \ldots \lamA_k$ are the $k$ largest
eigenvalues of the adjacency matrix, and if $d_1 \ge \ldots \ge d_k$ are the $k$ largest
degrees, then\footnote{These results are proven on a model with a slightly different generative procedure;
  we do not find that they yield a particularly good approximation for our experiments, that are
  conducted at the quite low $n=100$.}.
\begin{equation}
  \lamA_i = (1 + \o{1}) \sqrt{d_i} \quad \text{with high probability.}
\end{equation}
In \cite{Farkas2001}, the authors demonstrate numerically that the adjacency spectrum exhibits a
triangular peak with power-law tails.

Having a high degree makes a vertex more likely to attract more connections, so the graph quickly
develops strongly connected ``hubs,'' or vertices with very high degree, which cannot be found in the
\ER model. This impacts both the global and local structure of the graph. Hubs are by definition
global structures, as they touch a significant portion of the rest of the graph, making path lengths
shorter and increasing connectivity throughout the graph. On the local scale, vertices in the graph
tend to connect exclusively to the highest-degree vertices in the graph, rather than to one another,
generating a tree-like topology. This particular topology yields a signature in the tail of the
spectrum.
\subsubsection{The Watts-Strogatz model}
Many real-world graphs exhibit the so-called ``small world phenomenon,'' where the expected shortest
path length between two vertices chosen uniformly at random grows logarithmically with the size of
the graph. Watts and Strogatz \cite{Watts1998} constructed a random graph model that exhibits this
behavior, along with a high clustering coefficient not seen in an uncorrelated random graph. The
clustering coefficient is defined as the ratio of number of triangles to the number of connected
triplets of vertices in the graph. The \textbf{Watts-Strogatz model} \cite{Watts1998} is designed to
be the simplest random graph that has high local clustering and small average shortest path
distance between vertices.

Like the preferential attachment model, this graph is most easily described via a generative
mechanism. The algorithm proceeds as follows. Let $n$ be the size of the desired graph, let
$0\leq p \leq 1$, and let $k$ be an even integer, with $k<n$. We begin with a ring lattice, which is
a graph where each vertex is attached to its $k$ nearest neighbors, $k/2$ on each side. We then
randomly rewire edges (effectively creating shortcuts) as follows. With probability $p$, each edge
$(i,j)$ (where $i<j$) is replaced by the edge $(i,l)$, where $l$ is chosen uniformly at random. The
target $l$ is chosen so that $i\neq l$ and $i$ is not connected to $l$ at the time of rewiring. We
stop once all edges have been iterated through. We add an additional stipulation that the graph must
be connected. If the algorithm terminates with a disconnected graph, then we restart the algorithm
and generate a new graph.

As mentioned before, the topological features that are significant in this graph are the high local
clustering and short expected distance between vertices. Of course, these quantities are dependent
on the parameter $p$; as $p\rightarrow 1$, the Watts-Strogatz model approaches an uncorrelated
random graph. Similarly, as $p\rightarrow 1$ the adjacency spectral density transitions from the
tangle of sharp maxima typical of a ring-lattice graph to the smooth semi-circle of the uncorrelated
random graph \cite{Farkas2001}. Unlike the models above, this model exhibits primarily local
structure. Indeed, we observe that the most significant differences lie in the tail of the
adjacency spectrum, that can be directly linked to the number of triangles in the graph
\cite{Farkas2001}. On the large scale, however, this graph looks much like the uncorrelated random
graph, in which it exhibits no communities or high-degree vertices.

This model fails to produce the scale-free behavior observed in many real world networks. Although
the preferential attachment model reproduces this scale-free behavior, it fails to reproduce the
high local clustering that is frequently observed, and so we should think of neither model as fully
replicating the properties of observed graphs.
\subsubsection{The Configuration model}
The above three models are designed to mimic certain properties of real world networks. In some cases,
however, we may wish to create a random graph with a prescribed degree sequence. That is to say, we
seek a distribution that assigns equal probability to each graph, conditioned upon the graph having
a given degree sequence. The simplest model that attains this result is the configuration model
\cite{Bender1978}. Recently, Zhang et al. \cite{Zhang2014} have derived an asymptotic expression for
the adjacency spectrum of a configuration model, that is exact in the limit of large graph size and
large mean degree.

Inconveniently, this model is not guaranteed to generate a simple graph; the resulting graph can
have self-edges, or multiple edges between two vertices. In 2010, Bayati et al. \cite{Bayati2010}
described an algorithm that samples (approximately) uniformly from the space of simple graphs with a
given degree distribution. In \cite{Bayati2010} the authors prove that the distribution is
asymptotically uniform, but they do not prove results for finite graph size (see \cite{blitzstein11}
for a more detailed analysis). We use this algorithm despite the fact that it does not sample the
desired distribution in a truly uniform manner; the fact that the resulting graph is simple
overcomes this drawback.

We refer to graphs sampled in this way as {\em configuration model} graphs. The significance of this
class of graphs stems from the fact that we can use them to control for the degree sequence when
comparing graphs; they are used as a null model, similar to the uncorrelated random graph, but they
can be tuned to share some structure (notably, the power-law degree distribution of preferential
attachment) with the graphs to which they are compared.
\begin{figure}[H]
  \centerline{
    \includegraphics[width=0.7\textwidth]{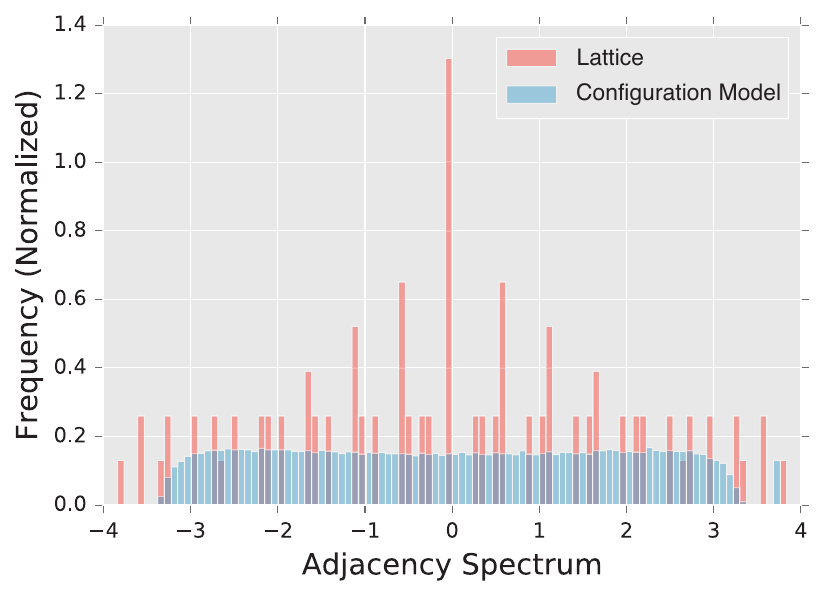}
  }
  \caption{Spectral densities $\lamA$ of the adjacency matrix for a lattice graph and a degree
    matched configuration model. Densities are built from an ensemble of 1,000 graphs generated using
    parameters described in Subsection\ref{sec:lattice_results}. 
    \label{fig:lattice_spectrum}}
\end{figure}
\subsubsection{Lattice graphs}
We use a 2-dimensional $x$ by $y$ rectangular lattice as a prototypical example of a highly regular
graph. This regularity is reflected by the discrete nature of the lattice's spectrum, which can be
seen in Fig \ref{fig:lattice_spectrum}. The planar structure of the lattice allows for an intuitive
understanding of the eigenvalues, as they approximate the vibrational frequencies of a
two-dimensional surface.

This is a particularly strong flavor of local structure, as it is not subject to the noise present in
random graph models. This aspect allows us to probe the functioning of our distances when they are
exposed to graphs with a high amount of inherent structure and very low noise.
\subsubsection{Exponential random graph models}
A popular random graph model is the \textbf{exponential random graph model}, or ERGM for
short. Although they are popular and enjoy simple interpretability, we do not use ERGMs in our
experiments. Unlike some of our other models that are described by their generative mechanisms,
these are described directly via the probability of observing a given graph $G$.

Let $g_i(G)$ be some scalar graph properties (e.g., size, volume, or number of triangles) and let
$\theta_i$ be corresponding coefficients, for $i=1,\ldots,K$. Then, the ERGM assigns to each graph a
probability \cite{Lusher2012}

\begin{equation*}
  \Prob{G} \propto \exp\left(\sum_{i=1}^K \theta_i g_i(G)\right).
\end{equation*}

This distribution can be sampled via a Gibbs sampling technique, a process that is outlined in
detail in \cite{Lusher2012}. ERGMs show great promise in terms of flexibility and interpretability;
one can seemingly tune the distribution towards or away from any given graph metric, including mean
clustering, average path length, or even decay of the degree distribution.

However, our experience attempting to utilize ERGMs led us away from this approach. When sampling
from ERGMs, we were unable to control properties individually to our satisfaction. We found that
attempts to increase the number of triangles in a graph increased the graph volume; when we
subsequently used the ERGM parameters to \textit{de-emphasize} graph volume, the sampled graphs had
an empirical distribution very similar to an uncorrelated random graph.
\subsubsection{Graph Neural Networks}
  Each one of the graph ensembles described in the previous sections represents the quintessential
  exemplar of a certain graph \emph{structure} (e.g, degree distribution, clustering coefficients,
  shortest path distance, community structure, etc.)  Each ensemble can be thought as the atomic
  building block that can be used to understand complex existing real world networks. For instance,
  it is shown in \cite{wolfe13} that any sufficiently large graph behaves approximately like a
  stochastic blockmodel. These networks are also amenable to a rigorous mathematical analysis, and
  one can analyze the influence on the graph distances of changes in the graph geometry and topology.

  As explained in section~\ref{DGK}, there has been some very recent attempts at generating random
  realizations of graphs by learning the structure of massive datasets of existing networks (e.g,
  \cite{li18,you18,shine19}). These algorithms offer an implicit representation of a set of graphs,
  by discovering an optimal neural network that can generate new graphs with similar structures. In
  contrast to the prototypical random graph ensembles, the current understanding of the theoretical
  properties of the graph neural networks is very limited: there are no results on the structural
  properties of these models (but see \cite{dehmamy19} for an estimate of the complexity of a graph
  convolutional network (number of nodes and number of hidden units) required to learn graph
  moments).

  A systematic study of the sensitivity of graph distances on graph neural networks is clearly
  needed. Such a study would provide information that would complement theoretical results that
  elucidate how expressive such graph models can be \cite{xu18,morris19}. Unfortunately, such
  experiments clearly go beyond the scope of the current manuscript.
\subsection{Real world networks
\label{sec:real_networks}}
Random graph models are often designed to simulate a single important feature of real world
networks, such as clustering in the Watts-Strogatz model or the high-degree vertices of the
preferential attachment model. In real networks, these factors coexist in an often unpredictable
configuration, along with significant amounts of noise. Although the above analysis of the efficacy
of various distances on random graph scenarios can help inform and guide our intuition, to truly
understand their utility we must also look at how they perform when applied to empirical graph data.

  \color{violet} In this study, we evaluate the performance of the aforementioned distances using two scenarios. First,
  we study the change point detection scenario for two time-varying networks: a dynamic social-contact graph, collected
  via RFID tags in an French primary school \cite{Stehle2011}, and a time series of emails exchanged between 986 members
  of a large European research institution \cite{leskovec07} over a period of 803 days.

Secondly, we investigate the two-sample test problem in neuroscience: given two populations of functional brain
connectivity networks, we compute a statistic to test whether both populations are generated by the same probability
distribution of controls (null hypothesis), or one population is significantly different from the other
one. Specifically, we compare the functional connectivity of subjects with a diagnosis of autism spectrum disorder
\cite{DiMartino2013} versus a population of controls.

\begin{figure}[H]
  \centerline{
    \includegraphics[width=\textwidth]{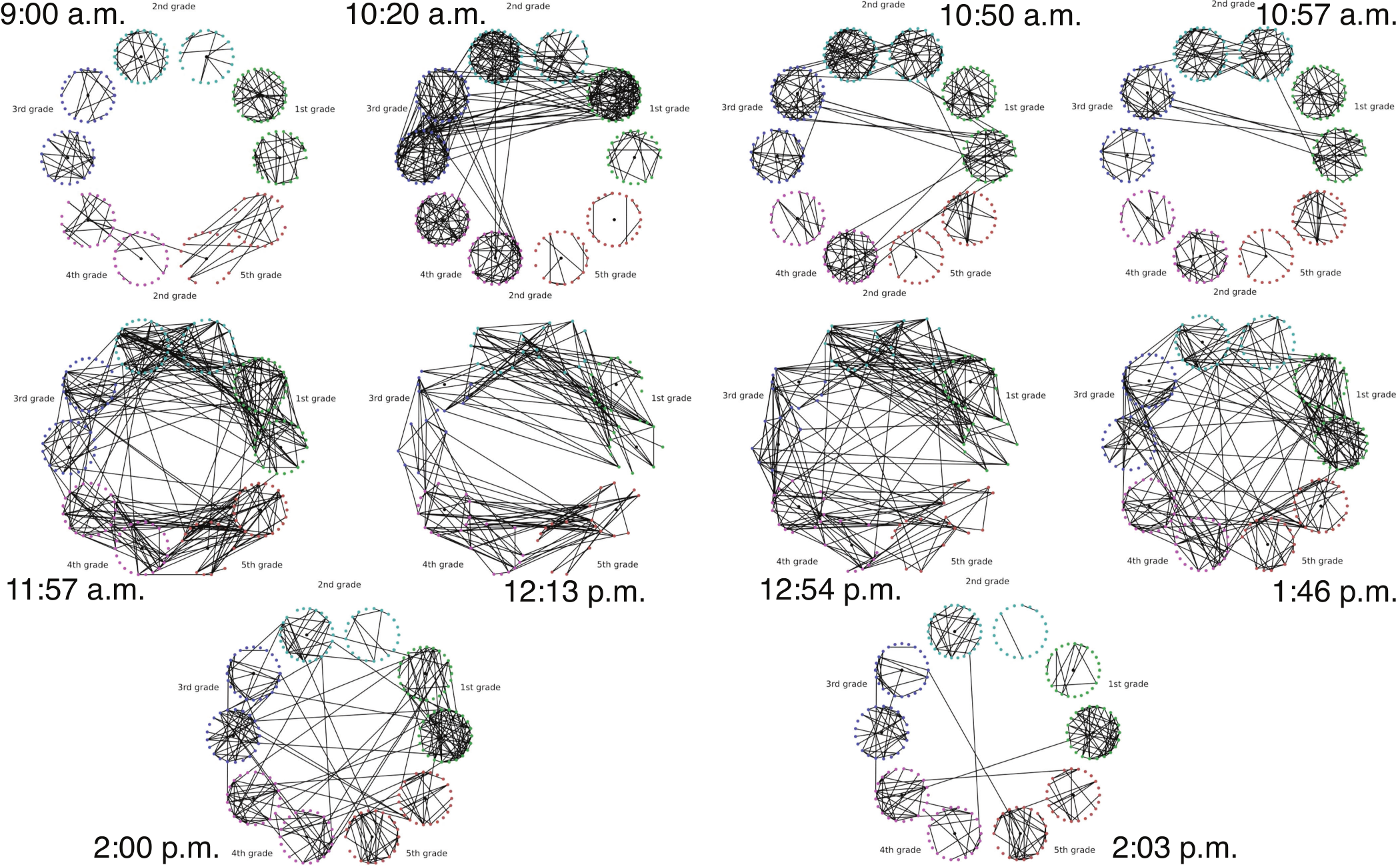}}
  \caption{Top to bottom, left to right: snapshots of the face-to-face contact network at times
    (shown next to each graph) surrounding significant topological changes.
    \label{fig:contact_graphs}}
\end{figure}
\subsubsection{Primary school face to face contact}
\label{sec:primary-school}
Some of the most well-known empirical network datasets reflect social connective
structure between individuals, often in online social network platforms such as
Facebook and Twitter. These networks exhibit structural features such as
communities and highly connected vertices, and can undergo significant
structural changes as they evolve in time. Examples of such structural changes
include the merging of communities, or the emergence of a single user as a
connective hub between disparate regions of the graph.
\paragraph{Description of the Experiment.}
The data are part of a study of face to face contact between primary school students~\cite{Stehle2011}. Briefly, RFID
tags were used to record face-to-face contact between students in a primary school in Lyon, France in October, 2009.
Events punctuate the school day of the children (see Table~\ref{tab:times}), and lead to fundamental topological changes
in the contact network (see Fig.~\ref{fig:contact_graphs}). The school is composed of ten classes: each of the five
grades (1 to 5) is divided into two classes (see Fig.~\ref{fig:contact_graphs}).

\begin{table}[H]
  \begin{center}
    \begin{small}
      \begin{tabular}{rl}
        \toprule
        Time & Event \\
        \midrule
        10:30 a.m. -- 11:00 a.m. &  Morning Recess \\
        12:00 p.m. -- 1:00 p.m. &  First Lunch Period\\
        1:00 p.m.  -- 2:00 p.m. &  Second Lunch Period \\
        3:30 p.m. -- 4:00 p.m. & Afternoon Recess\\
        \bottomrule
      \end{tabular}
    \end{small}
  \end{center}
  \caption{Events that punctuate the school day.
    \label{tab:times}}
\end{table}

The construction of a dynamic graph proceeds as follows: time series of edges that correspond to
face to face contact describe the dynamics of the pairwise interactions between students.  We divide
the school day into $N=150$ time intervals of $\Delta t \approx 200 \;\text{s}$. We denote by
$t_i = 0, \Delta t, \ldots, (N-1) \Delta t$, the corresponding temporal grid. For each $t_i$ we
construct an undirected unweighted graph $G_{t_i}$, where the $n=232$ nodes correspond to the $232$
students in the 10 classes, and an edge is present between two students $u$ and $v$ if they were in
contact (according to the RFID tags) during the time interval $[t_{i-1}, t_i)$.

For the purpose of this work, we think of each class as a community of connected
students; classes are weakly connected (e.g., see Fig.~\ref{fig:contact_graphs}
at times 9:00 a.m., and 2:03 p.m.). During the school day, events such as
lunchtime and recess, trigger significant increases in the the number of links
between the communities, and disrupt the community structure; see
Fig.~\ref{fig:contact_graphs} at times 11:57 a.m., and 1:46 p.m..
\color{violet}
\subsubsection{
European Union  Emails}
\color{violet}
\paragraph{Description of the Data.}
The data were obtained from the Stanford Large Network Dataset Collection \cite{snap}. The network was generated using
anonymized emails exchanged between 986 members of a large European research institution \cite{leskovec07}. There are
986 nodes that correspond to distinct individuals sending and receiving emails. To reduce the variability in the data,
we aggregate the emails exchanged every week, and perform an analysis at the week level. An edge was created between
nodes $i$ and $j$ if both $i$ sent at least one email to $j$ and $j$ sent at least one email to $i$ during that
week. The timeline starts on October 1, 2003 \cite{leskovec07}. The graph distances were computed between the weekly
graphs thus created.
\subsubsection{Functional brain connectivity
\label{sec:fmri-data}}
Graph theoretical analysis of the connective structure of the human brain is a popular research
topic, and has benefited from our growing ability to analyze network topology
\cite{sporns2004,sporns16,fornito16}. In these graph representations of the brain, the vertices are
physical regions of the brain, and the edges indicate the connectivity between two regions. The
connective structure of the brain is examined either at the ``structural'' level, in which edges
represent anatomical connection between two regions, or at the ``functional'' level, in which an
edge connects regions whose activation patterns are in some sense similar. Developmental and mental
disorders such as autism spectrum disorder \cite{Subbaraju2017} and schizophrenia \cite{Fornito2012}
have been shown to have structural correlates in the graph representations of the brains of those
affected. In this study we focus on autism spectrum disorder, or ASD.

\paragraph{Description of the Data.}
The Autism Brain Imagine Data Exchange \cite{abide,abide2}, or ABIDE, is an aggregation of
brain-imaging data sets from laboratories around the world that study ASD \cite{DiMartino2013}. The
data that we focus on are measurements of the activity level in various regions of the brain,
measured via functional magnetic resonance imaging (fMRI).

After preprocessing, the data is analyzed for quality. Of the original 1114 subjects (521 ASD and
593 TD), only 871 pass this quality-assurance step. These subjects are then spatially aggregated via
the Automated Anatomical Labelling (AAL) atlas, that aggregates the spatial data into 116 time
series.

To construct a graph from these time series, the pairwise Pearson correlation is calculated to
measure similarity. If we let $u$ and $v$ denote two regions in the AAL atlas and let $\rho(u,v)$
denote the Pearson correlation between the corresponding time series, the simplest way to build a
graph is to assign weights $w(u,v)=|\rho(u,v)|$. We exclude low correlations, as these are often
spurious and not informative as to the structure of the underlying network, and define the weights
\begin{equation*}
  w(u,v) = \left\{
    \begin{array}{ll}
      |\rho(u,v)| & \quad |\rho(u,v)| \geq T \\
      0 & \quad |\rho(u,v)| < T.
    \end{array}
  \right.
\end{equation*}
Finally, we also construct an unweighted graph according to
\begin{equation*}
  w(u,v) = \left\{
    \begin{array}{ll}
      1 & \quad |\rho(u,v)| \geq T \\
      0 & \quad |\rho(u,v)| < T.
    \end{array}
  \right.
\end{equation*}
We will compare both binary and weighted connectomes, generated for multiple thresholds. This will
allow us to be confident that our results are not artifacts of poorly chosen parameters in our
definition of the connectome graph.
\subsection{Evaluation protocol: the distance contrast
  \label{sec:how_to_measure}}
\subsubsection{The Distance contrast}
The experiments are designed to mimic a scenario in which a practitioner is
attempting to determine whether a given graph belongs to a population or is an
outlier relative to that population. 

Specifically, let us define by $\cG_0$ and $\cG_1$ two graph populations,which we  refer to
as the \textbf{null} and \textbf{alternative} populations respectively. For each distance measure,
let $\cD_0$ be the distribution of distances $d(G_0,G_0')$ where $G_0$ and $G_0'$ are both drawn
from the distribution $\cG_0$. Similarly, let $\cD_1$ be the distribution of distances $d(G_0,G_1)$,
where $G_0$ is drawn from $\cG_0$ and $G_1$ is drawn from $\cG_1$.

The statistic $\cD_0$ characterizes the natural variability of the graph population $\cG_0$, as seen
through the lens formed by the distance $d$. Similarly, the statistic $\cD_1$ reveals how
distant -- according to the distance $d$ -- the two graph populations $\cG_0$ and $\cG_1$ are.  If
the distributions of $\cD_0$ and $\cD_1$ are well separated, then $d$ is effective at
differentiating the null population from the alternative population.

To that end, we normalize the statistics of $\cD_1$ by those of $\cD_0$ in order to compare. In
particular, let $\mu_i$ be the sample mean of $\cD_i$, and let $\sigma_i$ be the sample standard
deviation, for $i\in \{0,1\}$. We define the following (normalized) contrast, $\hcD_1$,
\cite{Casella2002}, between $D_0$ and $D_1$, whose samples $\hD_1$ are calculated via
\begin{equation}
  \label{eq:4}
  \hD_1 \eqdef \frac{D_1 - \mu_0}{\sigma_0}.
\end{equation}
This studentized distance contrast can also be related to the Wald test statistic \cite{Casella2002}
\begin{equation}
  \label{eq:wald}
  \frac{D_0 - \mu_0}{\sigma_0}.
\end{equation}
If the empirical distribution of contrast $\hcD_1$ is well separated from zero, viz. the contrast
between $D_1$ and the sample mean $\mu_0$ is significantly greater than the standard deviation, then
the distance is effectively separating the null and alternative populations.
\subsubsection{Comparisons of the random graph ensembles}
Table \ref{tab:comparisons} describes the various experiments. Each model is compared against a null
model; the null model can be sampled either from the \ER model, or from a configuration model. The
latter makes it possible to match the degree distribution of the model being tested against that of
the null model. We compare the distance contrast in \eqref{eq:4} between each model and the null
model using all the distances. When appropriate, we also report the performance of the spectral
distances for various $k$. Table \ref{tab:comparisons} also displays the structural feature that is
being evaluated for a particular experiment.
\begin{table}[H]
  \begin{center}
    \begin{tabular}{|l|l|l|l|}
      \toprule
      \bf Section & \bf Null & \bf Alternative & \bf Structural Difference \\
      \midrule
      Stochastic Block Model & $G(n,p)$ & SBM & Community structure \\\hline
      Preferential Attachment & $G(n,p)$ & PA & High-degree vertices \\\hline
      Preferential Attachment vs & CM & PA & Structure not in the\\
      Configuration Model &  &  & degree sequence\\\hline
      Watts-Strogatz & $G(n,p)$ & WS & Local structure \\\hline
      Lattice Graph & $G(n,p)$ & Lattice & Extreme local structure\\\hline
    \end{tabular}
  \end{center}
  \caption{Table of comparisons performed, and the important structural features therein. $G(n,p)$
    indicates the \ER uncorrelated random graph, SBM is the stochastic blockmodel, PA is the
    preferential attachment model, CM is the degree matched configuration model, and WS is the
    Watts-Strogatz model.}
  \label{tab:comparisons}
\end{table}
\subsubsection{Comparisons of the real world networks}
\paragraph{Primary School Face to Face Contact and  European Union Emails.} Temporal changes in the graph topology over
time  are quantified using the various distance measures. For each distance measure $d$, we defined the
following temporal difference,
\begin{equation*}
  D_R(t_i) \eqdef d(G_{t_{i-1}},G_{t_i}).
\end{equation*}
To help compare these distances with one another, we normalize each by its sample mean
$\overline{D} = N^{-1}\sum_{i} D(t_i)$, and we define the normalized temporal difference,
\begin{equation*}
  \hD(t) = D(t) / \overline D.
\end{equation*}
\paragraph{Functional Brain Connectivity.} 
We define $\cG_1$ to be the set of connectomes computed from the ASD subjects, and $\cG_0$ the set
of connectomes from the control population (null model). The evaluation proceeds as described in
Section \ref{sec:how_to_measure}. The distance contrast between the two populations is evaluated
using the statistic defined in \eqref{eq:4},
\begin{equation}
  \hD_1 \eqdef \frac{D_1 - \mu_0}{\sigma_0}.
\end{equation}

\section*{Results}
\setcounter{section}{3}\setcounter{subsection}{0} \setcounter{subsubsection}{0}
\subsection{Random graph ensembles
  \label{sec:results_graph_models}}
  \color{violet}
  For each experiment described in Table \ref{tab:comparisons}, we generate 50 samples of $D_0$ and
$D_1$, where each sample compares two graphs of size $n=1,000$, unless otherwise specified.
The graphs
are always connected; the sampler will discard a draw from a random graph distribution if the
resulting graph is disconnected. Said another way, we draw from the distribution defined by the
model, conditioning upon the fact that the graph is connected.

The small size of the graphs allows us to use larger sample sizes; although all of the matrix distances
used have fast approximate algorithms available, we use the slower, often $\O{n^2}$, exact
algorithms for the  experiments, and so larger graphs would be prohibitively slow to work with. In
all the experiments, we choose  parameter values so that the expected volume of the two models
under comparison is equal.

We display the performance of the various distances on the same figure. Boxes extend from lower to
upper quartile, with center line at median. Whiskers extend from 5th to 95th percentile (e.g., see
Fig \ref{fig:SBM_results}). We also display the performance of the spectral distances contrast $\hD_1$
-- for the three matrices: adjacency, combinatorial Laplacian, and normalized Laplacian as a function
of the number of eigenvalues used to compute the distance (e.g., see Fig \ref{fig:SBM_spec})
\subsubsection{Stochastic blockmodel}
\label{sec:SBM_results}
\begin{figure}[H]
  \centerline{
    \includegraphics[width=0.7\textwidth]{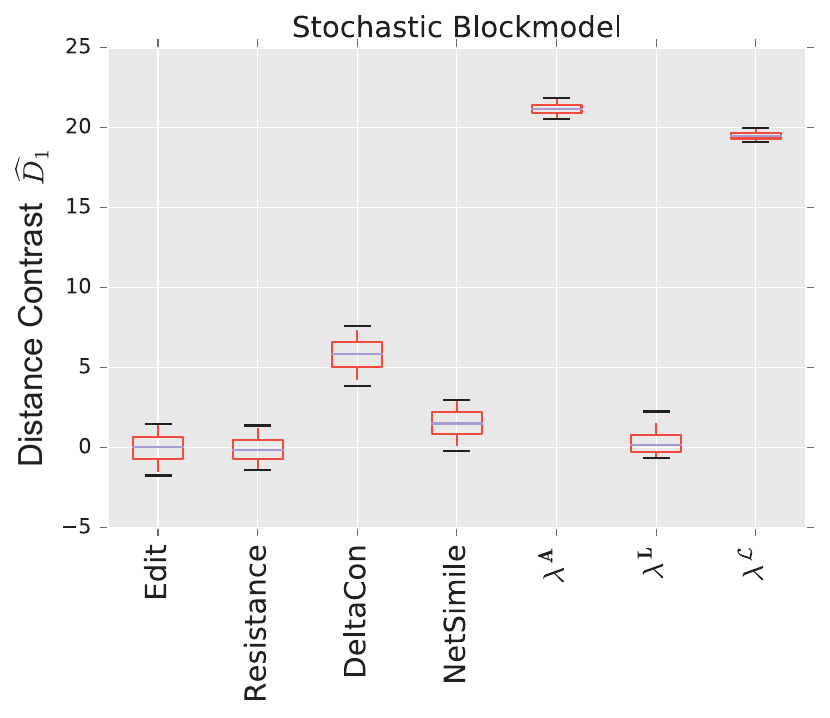}
  }
  \caption{\color{violet}
    Distance contrast $\hD_1$ between the stochastic blockmodel and the uncorrelated random
    graph model (null model).} 
  \label{fig:SBM_results}
\end{figure}

Fig~\ref{fig:SBM_results} displays the comparison between a stochastic blockmodel and an
uncorrelated random graph model (null model). The edge density $p=0.12$ of the uncorrelated random
graph is chosen so that graphs are connected with high probability\footnote{With
  these parameters, we observe that the empirical probability of generating a disconnected
  uncorrelated random graph with these parameters is $\sim 0.02\%$. The preferential attachment
  section describes in more detail why this exact value is chosen.}.

\color{violet}
  The stochastic blockmodel is composed of two communities of equal size, $n/2=500$.  Stochastic blockmodel experiments
  are run with in-community parameter $p = 1.9 \times 0.02$, and cross-community parameter $q = 0.1 \times 0.02$. Thus, the
  in-community connectivity is denser than the cross-community connectivity by a factor of $p/q=19$.
  \color{violet}

Since we have matched the volume of the graphs, the edit distance fails to distinguish the two
models. Among the matrix distances, \DC separates the two models most reliably. The adjacency and
normalized Laplacian distances perform well. The resistance perturbation distance and the
non-normalized Laplacian distance fail to distinguish the two models.

\begin{figure}[H]
  \centerline{
    \includegraphics[width=\textwidth]{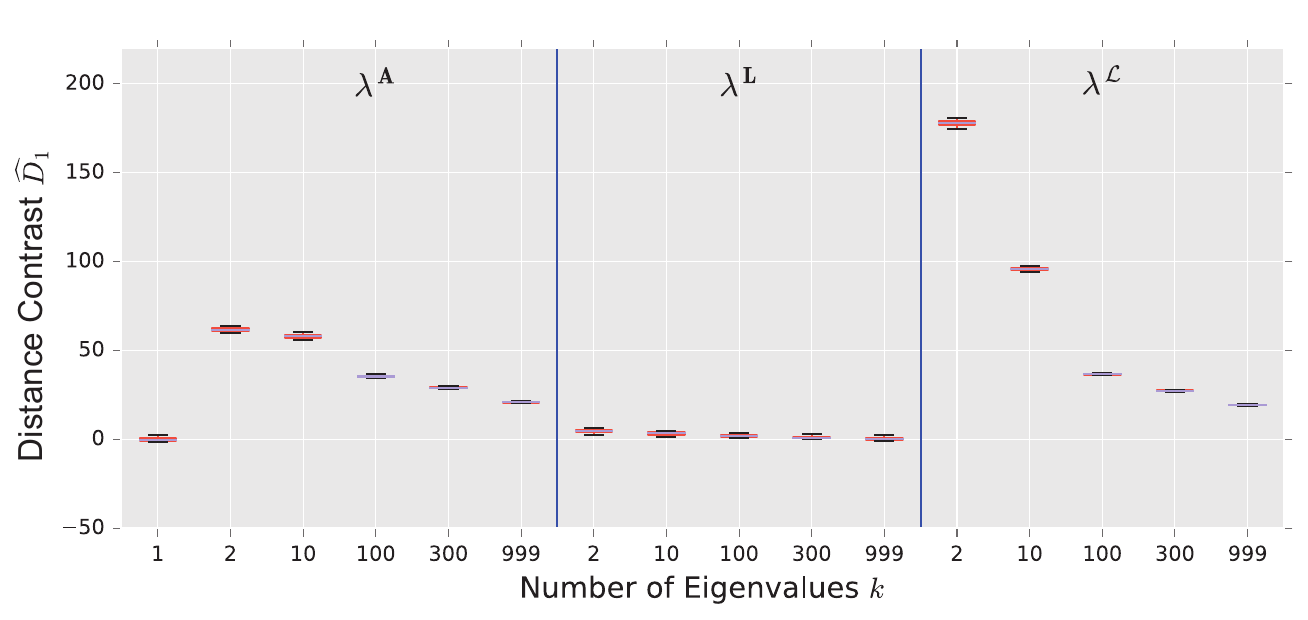}
  }
  \caption{\color{violet}
Spectral distances contrast $\hD_1$ -- for the three matrices: adjacency, combinatorial
    Laplacian, and normalized Laplacian (from left to right) -- between the stochastic blockmodel and
    the uncorrelated random graph model (null model). 
  \label{fig:SBM_spec}}
\end{figure}

As confirmed in Fig~\ref{fig:SBM_spec}, the performance of the adjacency distance is primarily driven by
differences in the second eigenvalue $\lamA_2$, and including further eigenvalues adds no benefit;
the normalized Laplacian also shows most of its benefit in the second eigenvalue $\lamNL_2$, but
unlike the adjacency distance, including more eigenvalues decreases the performance of the metric.


\subsubsection{Preferential attachment vs uncorrelated}
\label{sec:PA_results}
\begin{figure}[H]
  \centerline{
    \includegraphics[width=0.7\textwidth]{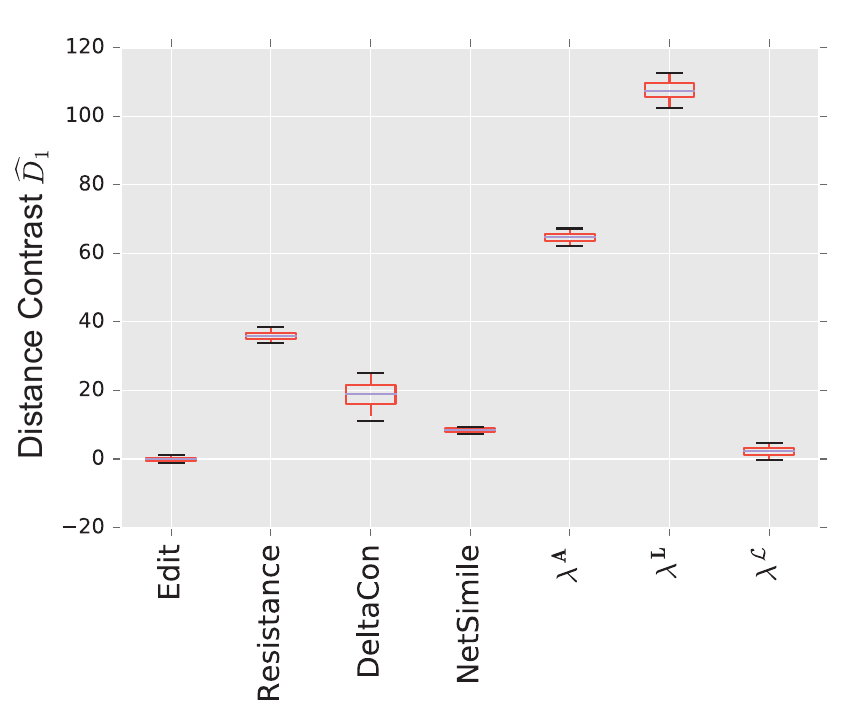}
  }
  \caption{
Distance contrast $\hD_1$ between the preferential attachment  and the uncorrelated random
    graph model (null model).
  \label{fig:PA_results}}
\end{figure}

Fig~\ref{fig:PA_results} shows the results of comparing a preferential attachment graph to an
uncorrelated random graph. The preferential attachment graph is quite dense, with $l=6$. Since the
number of edges in this model is always $|E|=l(n-l)$, we determine the parameter $p$ for the
uncorrelated graph via
\begin{equation*}
  p(l) = l(n-l){n \choose 2}^{-1},
\end{equation*}
to guarantee that both graphs always have the same volume. 

Again, because both graphs have the same number of edges (with high probability), the edit distance
fails to distinguish the two models. The resistance distance shows mediocre performance, although 0
is outside the 95\% confidence interval. \DC exhibits extremely high variability, although it has
the highest median of the matrix distances.

\begin{figure}[H]
  \centerline{
    \includegraphics[width=\textwidth]{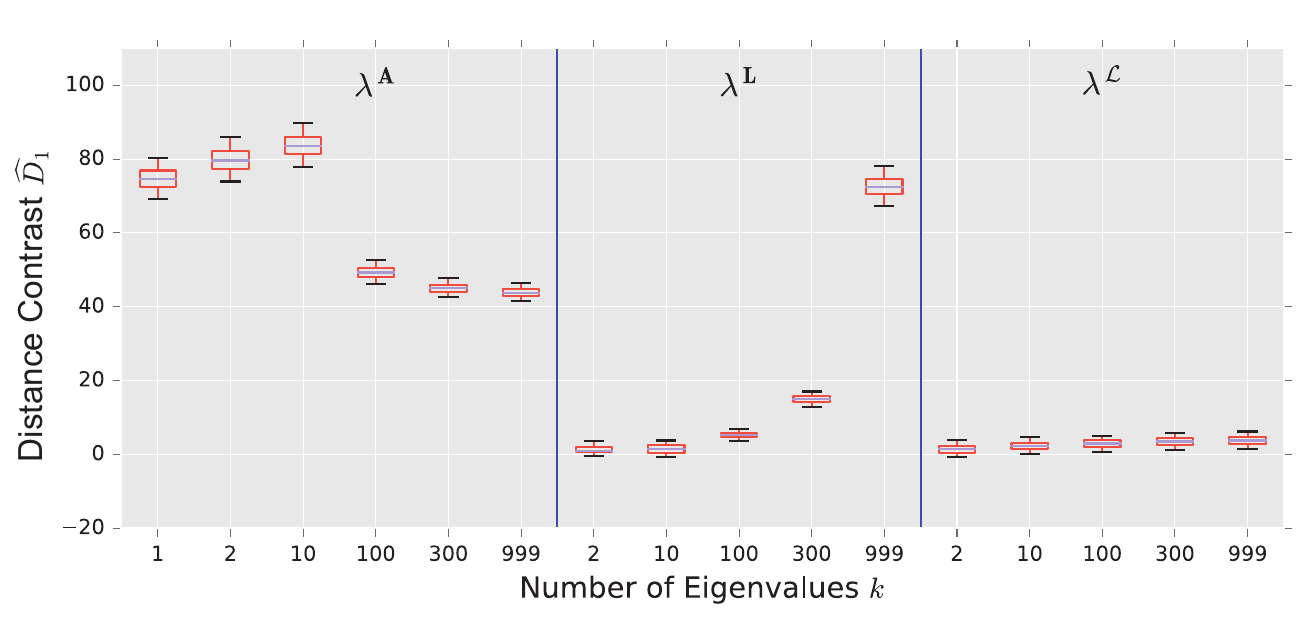}
  }
  \caption{\color{violet}
Spectral distance contrast $\hD_1$ -- for the three matrices: adjacency, combinatorial
    Laplacian, and normalized Laplacian (from left to right) -- between the preferential attachment
    and the uncorrelated random graph model (null model).  
    \label{fig:PA_k_results}}
\end{figure}

The combinatorial Laplacian distance outperforms all others, while the normalized Laplacian does not
separate the two models at all. Fig~\ref{fig:PA_k_results} shows that the very fine scale
eigenvalues of the combinatorial Laplacian (large index) are needed to differentiate the two
models. Conversely, the discriminating eigenvalues of the adjacency matrix are the smallest
eigenvalue; in fact, the first eigenvalue captures much of the contrast: the distance contrast
\eqref{eq:4} stays more or less constant as one increases $k$ (see Fig \ref{fig:PA_k_results}).


\subsubsection{Preferential attachment vs configuration model}
\label{sec:DD_PA_results}
To further explore the distinctive features of the preferential attachment graphs, we change here
the null model. Instead of using a volume matched uncorrelated random graph model, we use a degree
matched configuration model as the null model (the volume is automatically matched, since the number
of edges is half of the sum of the degrees). This experiment allows us to search for structure in
the preferential attachment model that is \emph{not} prescribed by the degree distribution.

\begin{figure}[H]
  \centerline{
    \includegraphics[width=0.7\textwidth]{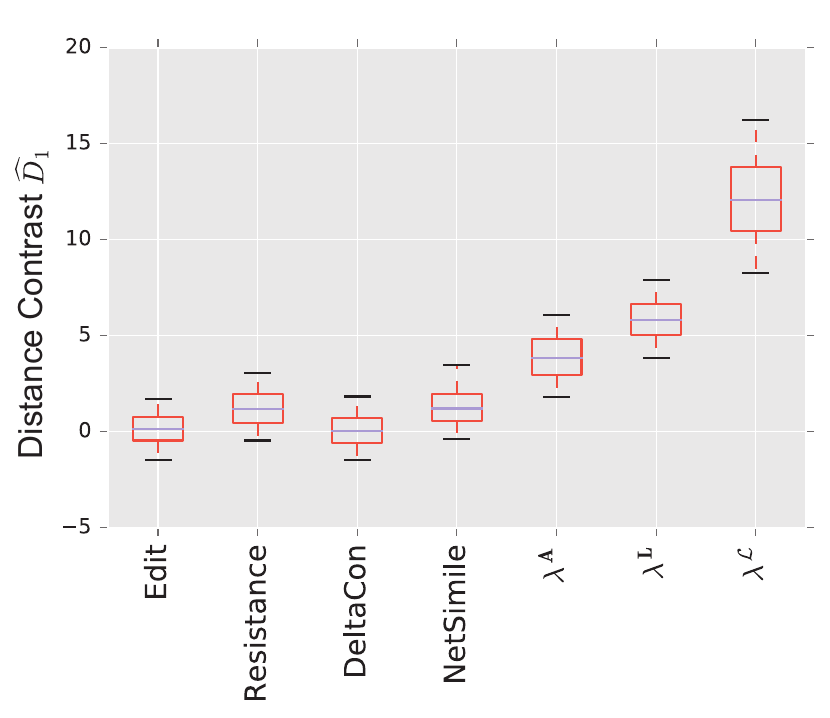}
  }
  \caption{Distance contrast $\hD_1$ between the preferential attachment model and the degree
    matched configuration model (null model).\label{fig:DD_PA_results}}
\end{figure}

An intriguing result happens: not a single distance can differentiate between a preferential
attachment graph and a randomized graph with the same degree distribution (see Fig
\ref{fig:DD_PA_results}).

The spectral distance based on the eigenvalues combinatorial Laplacian $\lamL$, which
yields the strongest contrast when comparing the preferential attachment model to the uncorrelated
random graph model is now unavailing. This thought-provoking experiment suggests that all
significant structural features of the preferential attachment model are prescribed by the degree
distribution.
\subsubsection{Watts-Strogatz}
\label{sec:WS_results}
  \color{violet}
The Watts-Strogatz experiments used $k = 20$, $p = 0.020020...$ (calculated so that
the volumes match) and $\beta = 0.1$. The number of nodes was $n=1,000$.

The Watts-Strogatz model is sparse, and thus the uncorrelated random graph has a low value of $p$ --
since we match the number of edges -- and is very likely disconnected. This is only a
significant problem for the resistance distance, that is undefined for disconnected graphs. To
remedy this, we use an extension of the resistance distance called the \textbf{renormalized
  resistance distance}, that is developed and analyzed in \cite{Wills2017}. This is the only
experiment in which the use of this particular variant of the resistance distance is required.

\begin{figure}[H]
  \centerline{
    \includegraphics[width=0.7\textwidth]{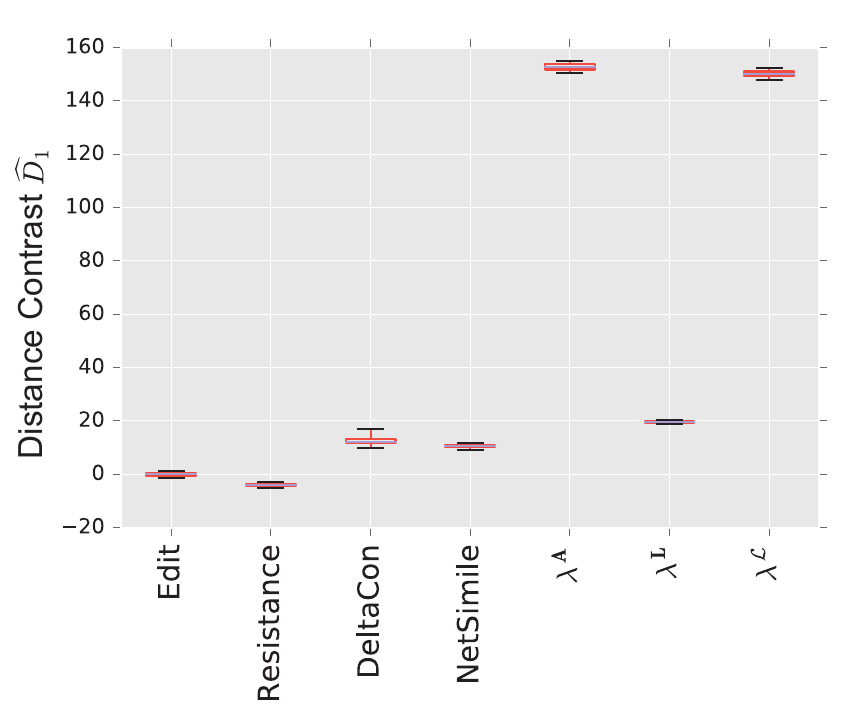}
  }
  \caption{
Distance contrast $\hD_1$ between a small-world graph  and the  the degree
    matched configuration model (null model). 
    \label{fig:WS_results}}
\end{figure}

Fig~\ref{fig:WS_results} shows that the spectral distances based on the adjacency and normalized
Laplacian are the strongest performers. Amongst the matrix distances, \DC strongly outperforms the
resistance distance. The resistance distance here shows a negative median, that indicates smaller
distances between populations than within the null population. This is likely due to the existence
of many (randomly partitioned) disconnected components within this particular null model, that
inflates the distances generated by the renormalized resistance distance. It is notable that,
contrary to the comparison in Section~\ref{sec:PA_results}, the normalized Laplacian outperforms the
combinatorial Laplacian.

\begin{figure}[H]
  \centerline{
    \includegraphics[width=\textwidth]{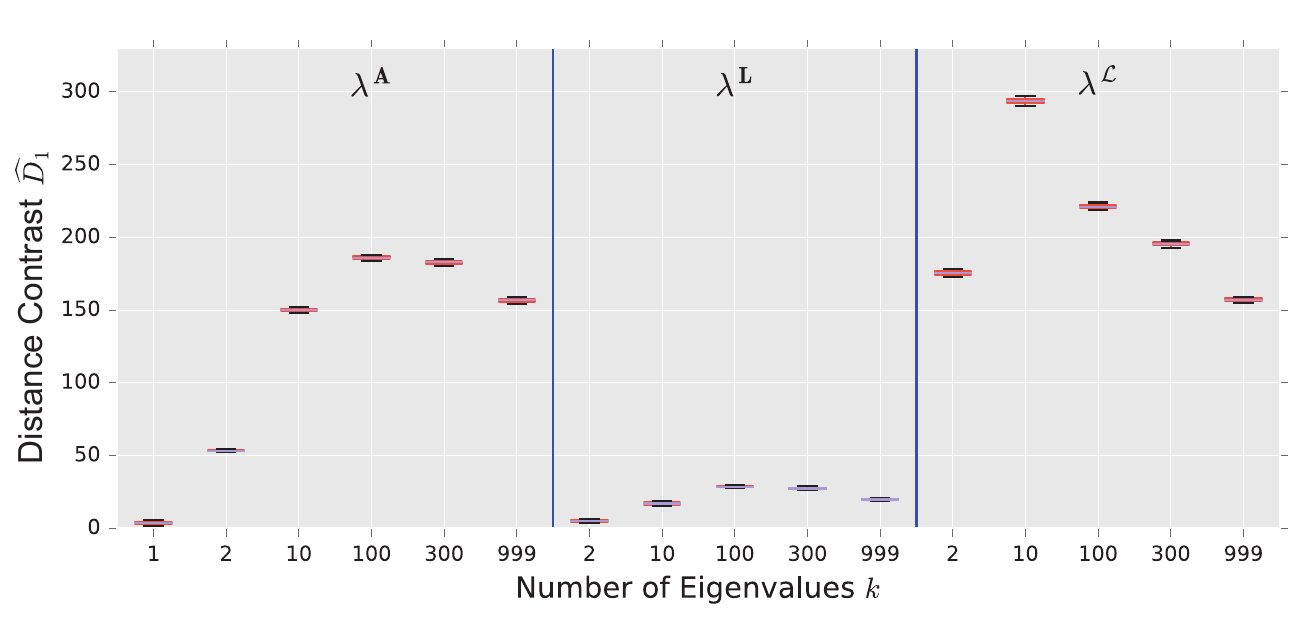}
  }
  \caption{
    Spectral diastances contrast $\hD_1$ -- for the three matrices: adjacency, combinatorial
    Laplacian, and normalized Laplacian (from left to right) -- between the small-world graph and the degree
    matched configuration model (null model). 
    \label{fig:WS_k_results}}
\end{figure}

Fig~\ref{fig:WS_k_results} displays the results for spectral distances, for a wide variety of
$k$. This figure is significant because it illustrates the fact that both coarse scales (large $k$
for $\lamA_k$) and fine scales (large $k$ for $\lamNL_k$) are necessary to yield the optimal 
contrast between the two models.

\subsubsection{Lattice graph}
\label{sec:lattice_results}
The final experiment, compares a lattice graph to a configuration model graph with the same degree
distribution.  The lattice graphs are $100 \times 10$, giving a total size of $1,000$.

\begin{figure}[H]
  \centerline{
    \includegraphics[width=0.7\textwidth]{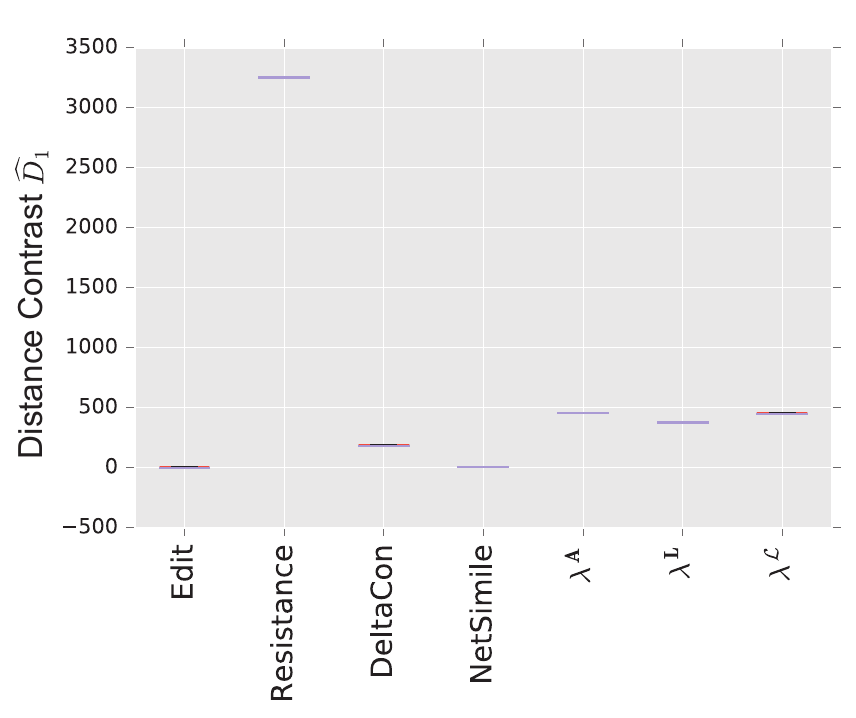}
  }
  \caption{\color{violet}
    Distance contrast $\hD_1$ between the $10 \times 100$ two-dimensional lattice graph and the the degree
    matched configuration model (null model).
    \label{fig:lattice_results}}
\end{figure}

The lattice here is highly structured, while the configuration model graph is quite
similar to an uncorrelated random graph; both the deterministic degree distribution of the lattice
and the binomial distribution of the uncorrelated random graph are highly concentrated around their
respective means.

We see in Fig~\ref{fig:lattice_results} that the scaled distances in this experiment are about an
order of magnitude higher than they are in other experiments for some of the distances; because the
lattice is such an extreme example of regularity, it is quite easy for many of the distances to
discern between these two models. The resistance distance has the highest performance, while
spectral distances all perform equally well. Note that for a regular graph, the eigenvalues of
$\bA$, $\bL$, and $\bcL$ are all equivalent, up to an overall scaling and shift, so we would expect
near-identical performance for graphs that are nearly regular.

\begin{figure}[H]
  \centerline{
    \includegraphics[width=\textwidth]{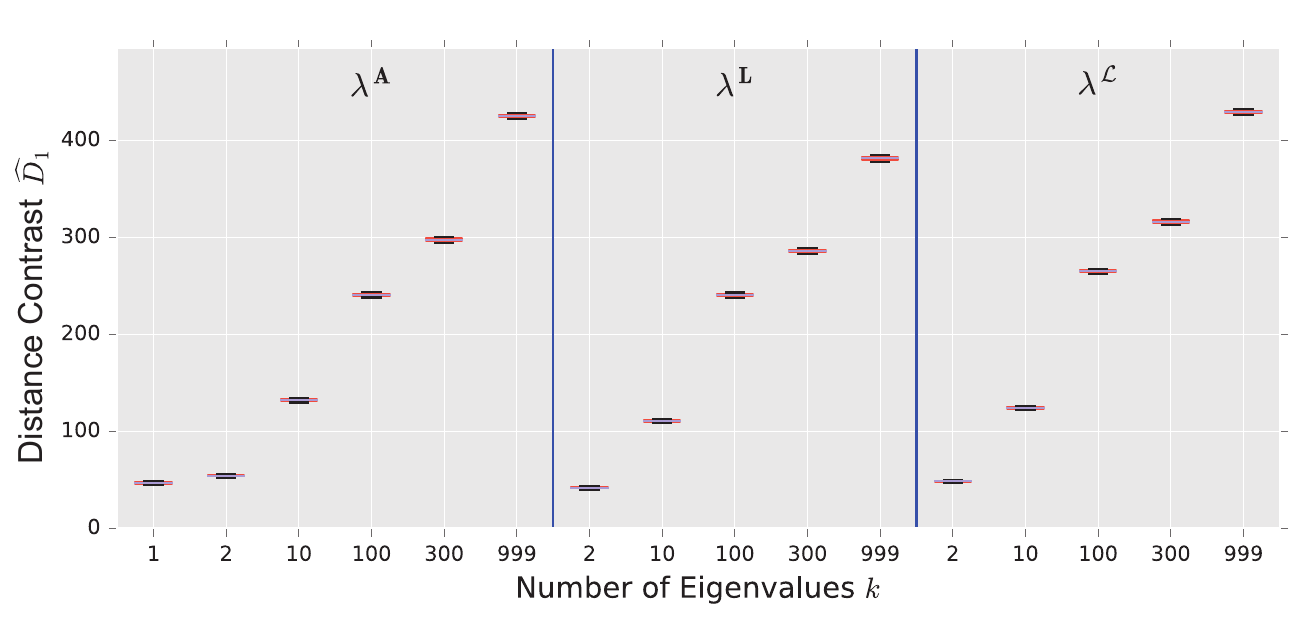}
  }
  \caption{\color{violet}
Spectral distances contrast $\hD_1$ -- for the three matrices: adjacency, combinatorial
    Laplacian, and normalized Laplacian (from left to right) --  between the  $10 \times 10$
    two-dimensional lattice graph and the degree matched configuration model (null model). 
    \label{fig:lattice_k_results}}
\end{figure}

The spectral distances need all the scales (i.e. all the eigenvalues) to discern between the lattice
and the configuration models (see Fig~\ref{fig:lattice_k_results}). This phenomenon, which is
similar to the Watts-Strogatz model (see Section~\ref{sec:WS_results}), points to the importance of the
local structure in the topology of the lattice graph

  \color{violet}

\subsection{Real world networks
\label{sec:results_real_networks}}
\subsubsection{Primary school face to face contact,
\label{sec:primary_xp}}
\begin{figure}[H]
  \centerline{
    \includegraphics[width=\textwidth]{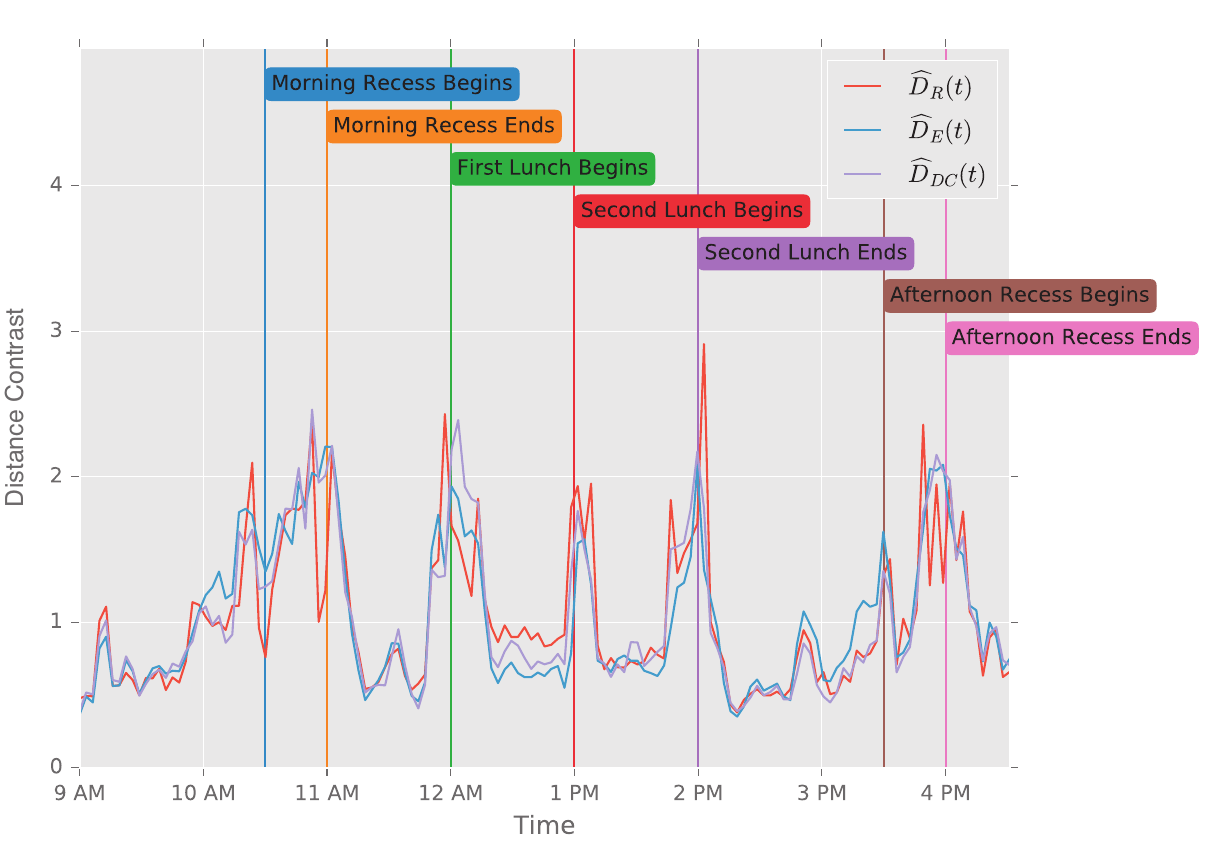}
  }
  \caption{Primary school data set: normalized temporal differences for the resistance distance
    $\hD_R$, edit distance $\hD_E$, and \DC distance $\hD_{DC}$.
    \label{fig:primary_school1}}
\end{figure}

Fig~\ref{fig:primary_school1} displays the normalized temporal differences for the resistance
distance $\hD_R$, edit distance $\hD_E$, and \DC distance $\hD_{DC}$. All the matrix distances are
capable of detecting significant changes in the hidden events that control the topology of the
contact network during the school day (see Fig \ref{fig:contact_graphs}). Indeed, the main
structural changes that the graph undergoes are transitions into and out of a strong ten-community
structure that reflects the classrooms of the school. For example, the adjacency matrix begins as
(mostly) block-diagonal at 9 AM, but has significant off-diagonal elements by morning recess at
10:20 AM, and is no longer (block) diagonally dominant come the lunch period at 12 PM.

\begin{figure}[H]
  \centerline{
    \includegraphics[width=0.8\textwidth]{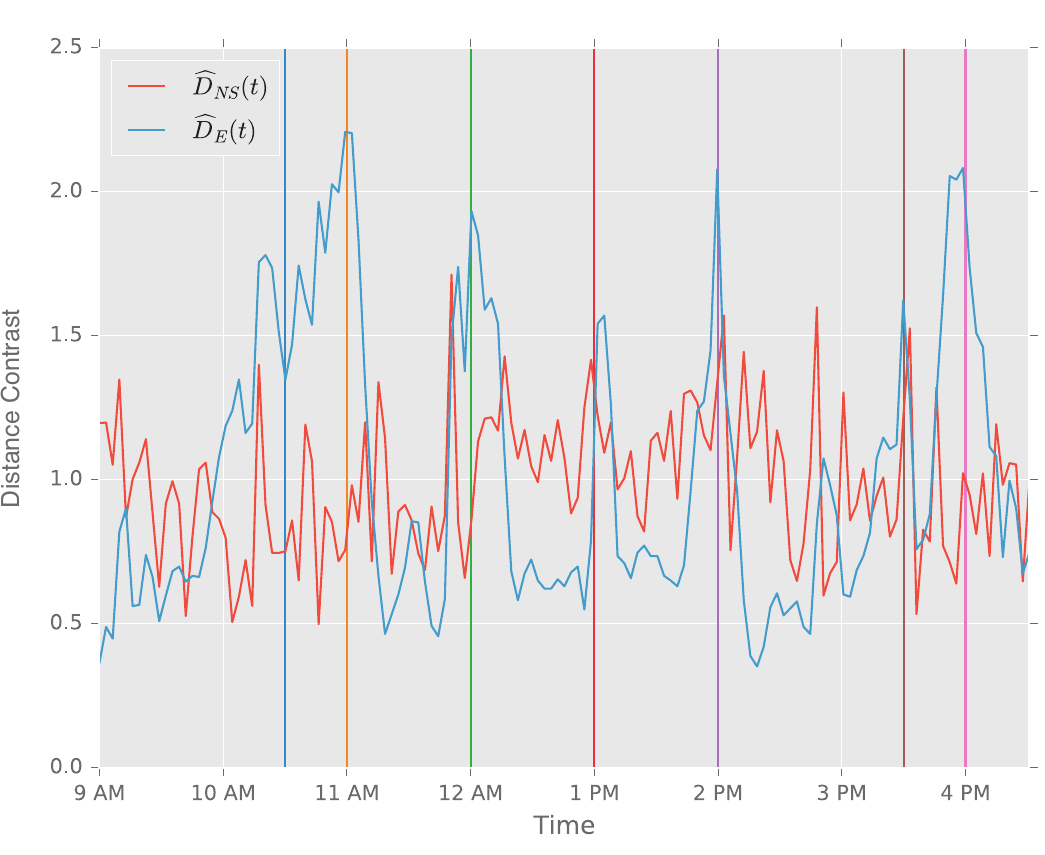}
  }
  \caption{Primary school data set: normalized temporal differences for the \NS distance $\hD_{NS}$
    and edit distance $\hD_E$.
    \label{fig:primary_school2}}
\end{figure}

There exists a persistent random variability of the very fine scale connectivity (e.g., edges come
and go within a community) that is superimposed on the large scale structural changes. Unlike, the
matrix distances (displayed in Fig~\ref{fig:primary_school1}), \textsc{NetSimile} is significantly
affected by these random fluctuations  (see Fig~\ref{fig:primary_school2}).

\begin{figure}[H]
  \centerline{
    \includegraphics[width=0.8\textwidth]{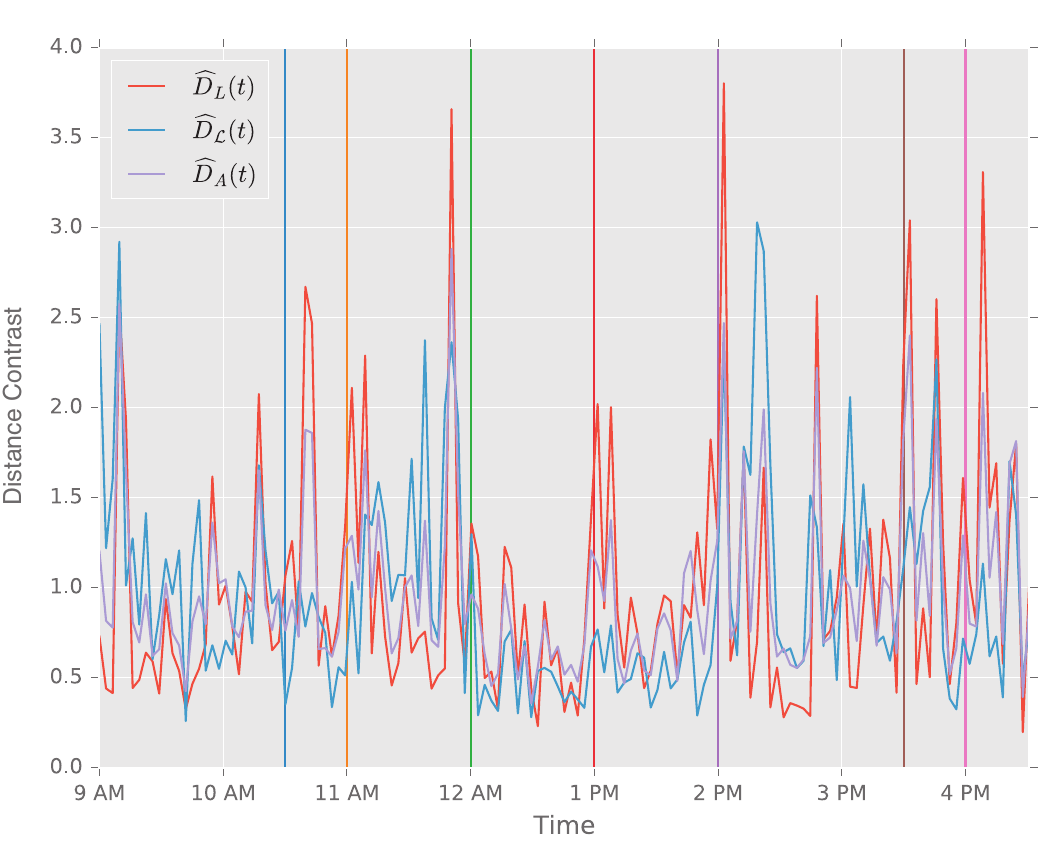}
  }
  \caption{Primary school data set: normalized temporal differences for the three spectral distances:
    combinatorial  Laplacian $\hD_L$, normalized Laplacian $\hD_{\mathcal{L}}$, and  adjacency $\hD_A$.
    \label{fig:primary_school3}}
\end{figure}

The stochastic variability in the connectivity appreciably influence the high frequency (fine scale)
eigenvalues. Consequently, the spectral distances, which are computed using all the eigenvalues,
lead to very noisy normalized temporal differences (see Fig~\ref{fig:primary_school3}), making it
difficult to detect the significant changes in the graph topology triggered by the school schedule.

  \color{violet}
  \subsubsection{European Union Emails
  \label{sec:email_xp}}
\begin{figure}[H]
  \centerline{
    \includegraphics[width=\textwidth]{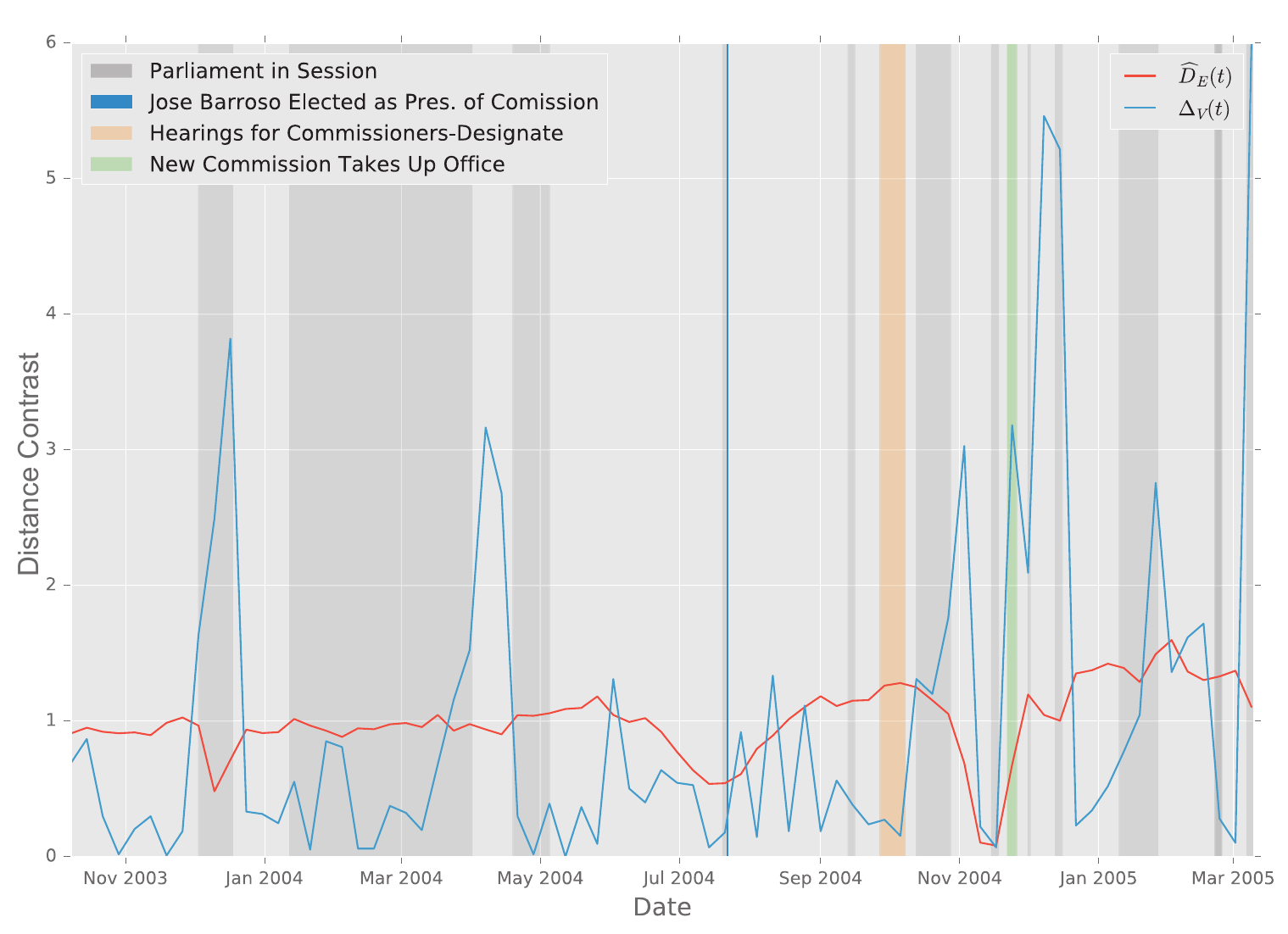}
  }
  \caption{\color{violet}
    EU-emails:  normalized temporal difference, for the edit distance $\hD_E$, and absolute value of the changes in the
    graph volume over time. 
    \label{fig:eu-emails-vol}}
\end{figure}

Fig. \ref{fig:eu-emails-vol} displays the changes in the volume of subsequent graphs (difference in the total number of
symmetric emails between two weeks), along with the edit distance, as a function of time. Inspired by the analysis of
the dataset performed by \cite{hajij18}, we superimposed some events that are related to the activity of the European
Parliament. These events were retrieved from \cite{europe}, and are displayed in Table \ref{table_EU}.\\

\begin{table}[H]
  \centering
  \begin{tabular}{ll}
    \toprule
    \color{violet} Date & \color{violet} Event\\
    \midrule
    \color{violet} 22 July & \color{violet} Jose Barroso elected as President of the Commission\\
    \color{violet}    27 September -- & \color{violet} Hearings of the Commissioners-designates, nominated \\
    \color{violet}  \hspace*{0.2pc}8 October     & \color{violet} by Jose Barroso\\
    \color{violet} 22 -- 26 November & \color{violet} New Commission takes up office\\
    \bottomrule
  \end{tabular}
  \caption{\color{violet}
        Events related to the activity of the European Parliament during 2004 \cite{europe}
    \label{table_EU}}
\end{table}

\begin{figure}[H]
  \centerline{
    \includegraphics[width=\textwidth]{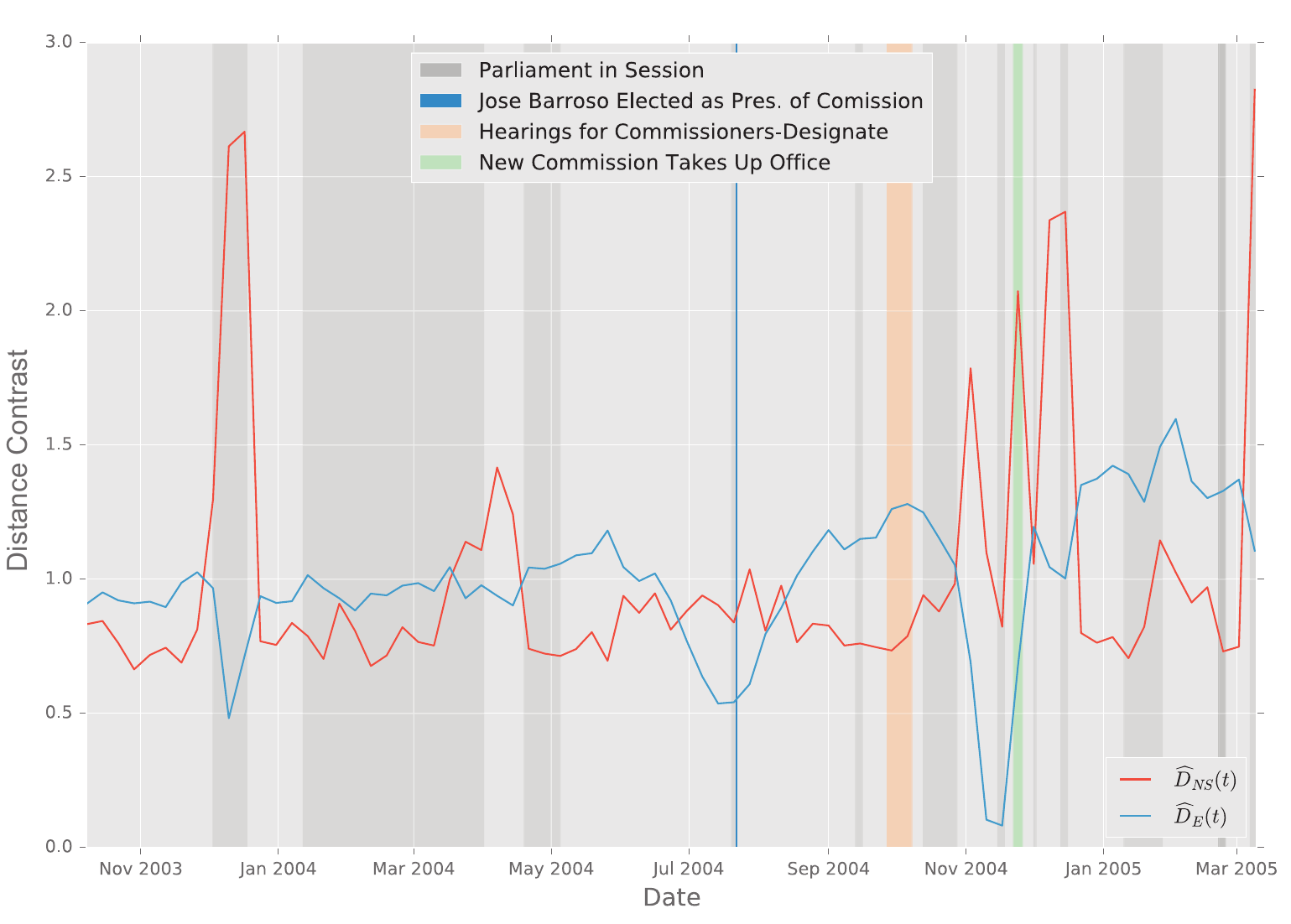}
  }
  \caption{\color{violet}
    EU-emails: normalized temporal differences for the \NS distance $\hD_{NS}$ and edit distance $\hD_E$. 
   \label{fig:eu-emails-NS}}
\end{figure}

\begin{figure}[H]
  \centerline{
    \includegraphics[width=\textwidth]{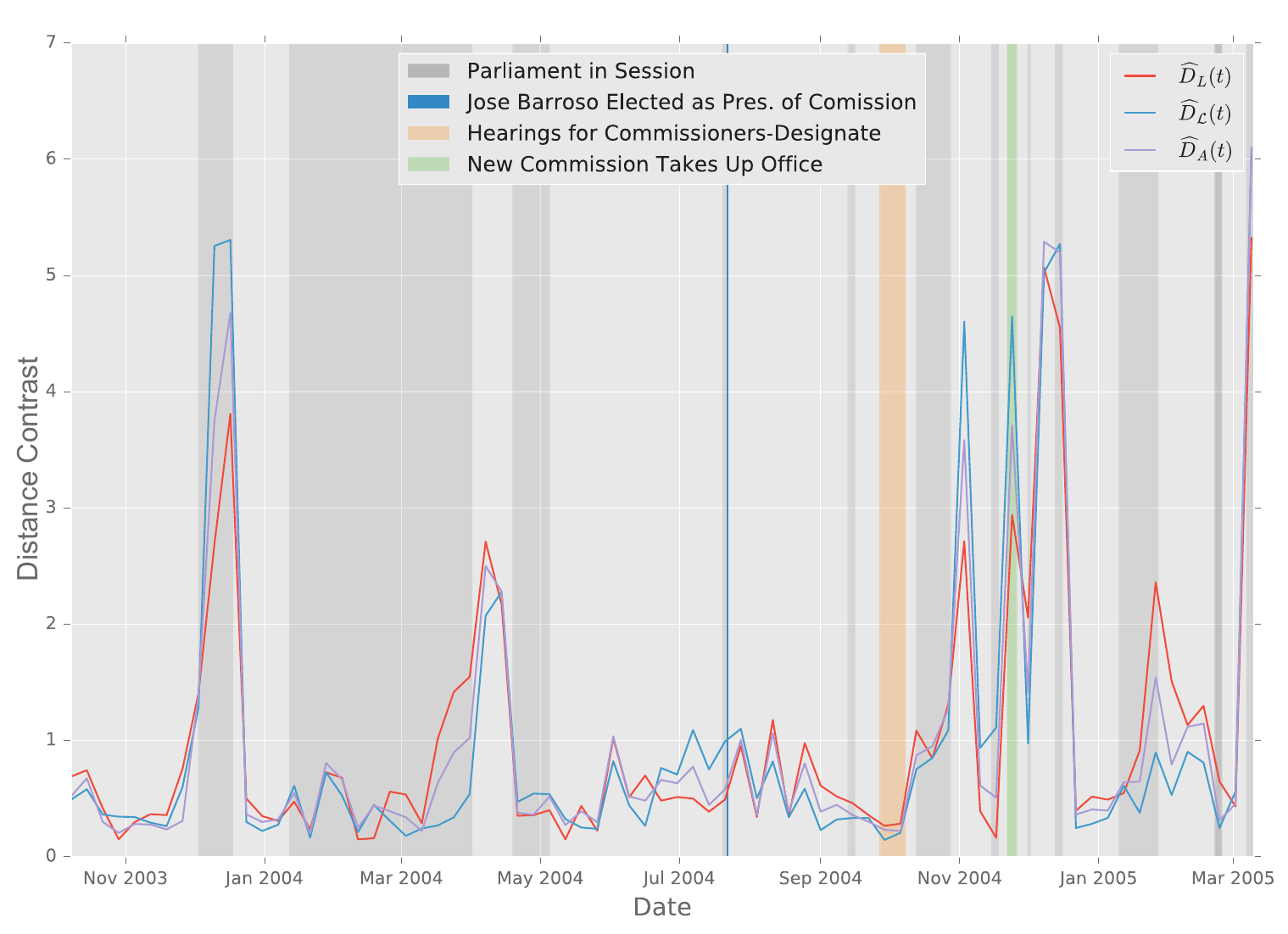}
  }
  \caption{\color{violet}
    EU-emails:  normalized temporal differences for the three spectral distances:
    combinatorial  Laplacian $\hD_L$, normalized Laplacian $\hD_{\mathcal{L}}$, and  adjacency $\hD_A$.  
    \label{fig:eu-emails-spectral}}
\end{figure}

\begin{figure}[H]
  \centerline{
    \includegraphics[width=\textwidth]{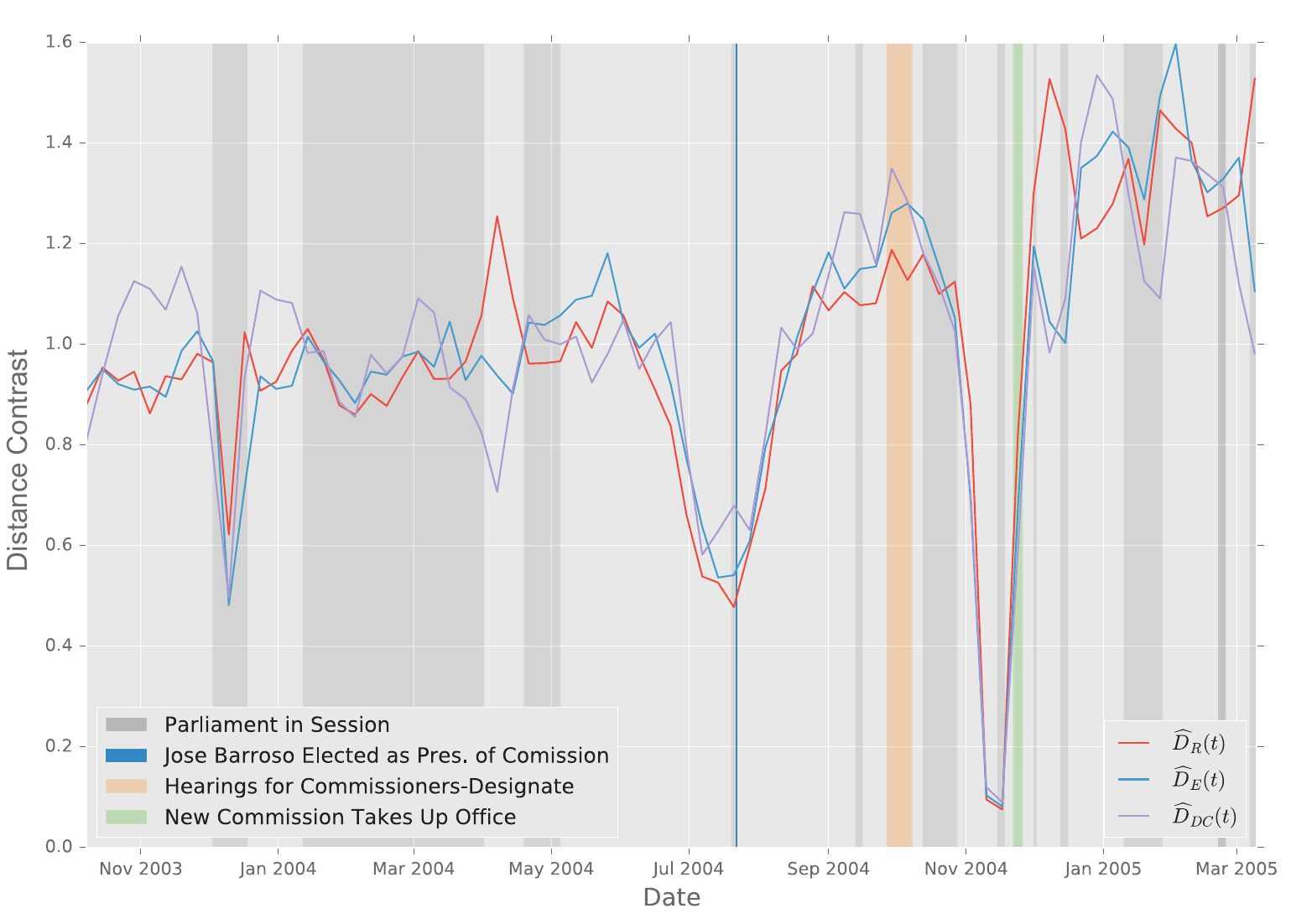}
  }
  \caption{\color{violet}
    EU-emails: normalized temporal differences for the resistance distance $\hD_R$, edit distance $\hD_E$, and \DC
    distance $\hD_{DC}$. 
    \label{fig:eu-emails-mat}}
\end{figure}

Fig.~\ref{fig:eu-emails-NS} displays the normalized temporal differences for the \NS distance $\hD_{NS}$and the edit distance
$\hD_E$.  Fig.~\ref{fig:eu-emails-spectral} displays the normalized temporal differences for three spectral distances:
combinatorial Laplacian $\hD_L$, normalized Laplacian $\hD_{\mathcal{L}}$, and adjacency $\hD_A$.  All the spectral
distances are correlated to the activity of the Parliament, including the hearings sessions and the entry into office of
the new 2004-2009 Commission, at the end of November 2004. Fig. \ref{fig:eu-emails-mat} displays the normalized temporal
differences for the resistance distance $\hD_R$, edit distance $\hD_E$, and \DC distance $\hD_{DC}$. All the matrix
distances are capable of detecting the election of Jose Barroso as President of the Commission, as well as the
investiture procedure of the 2004-2009 Commission: hearings in October 2004, and entry into office at the end of
November 2004.

\subsubsection{Functional brain connectivity
\label{sec:fmri_results}}
As explained in Section~\ref{sec:fmri-data}, the comparisons between sets of connectomes lead to two
types of analysis: the analysis of weighted (by the strength of the functional coupling between
brain regions) connectomes, and the comparison of unweighted (we only record if two regions are
functionally coupled -- irrespective of the strength of that coupling) connectomes. Furthermore, for
each type of connectome, we can vary the density of edges (and therefore the volume) by varying the
threshold used to define functional connectivity. We used two values of the threshold for the
Pearson correlation coefficient: 0.5 and 0.8.

\begin{figure}[H]
  \centerline{
    \includegraphics[width=0.7\textwidth]{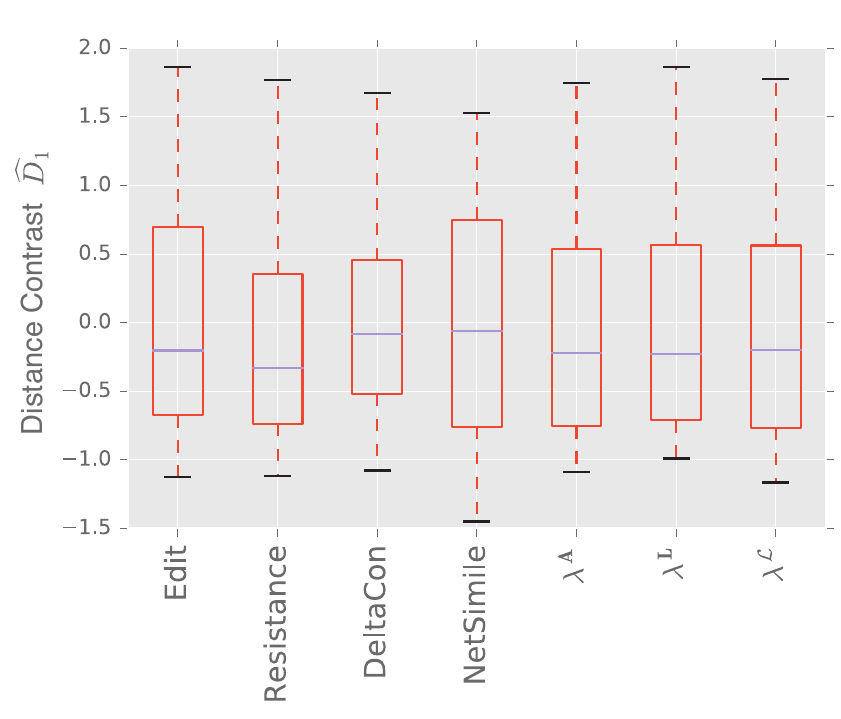}
  }
  \caption{ABIDE data set: distance contrast $\hD_1$ between the unweighted
    ASD and control connectomes for a threshold $T=0.5$
    \label{fig:ABIDE1}}
\end{figure}

\begin{figure}[H]
  \centerline{
    \includegraphics[width=0.7\textwidth]{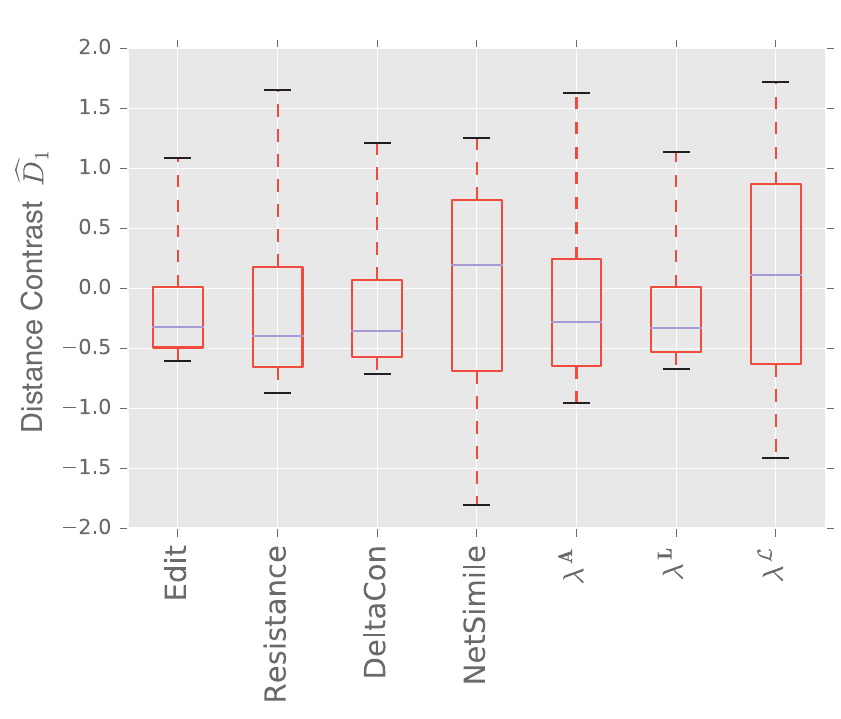}
  }
  \caption{ABIDE data set: distance contrast $\hD_1$ between the unweighted
    ASD and control connectomes for a threshold $T=0.8$
    \label{fig:ABIDE2}}
\end{figure}

\begin{figure}[H]
  \centerline{
    \includegraphics[width=0.7\textwidth]{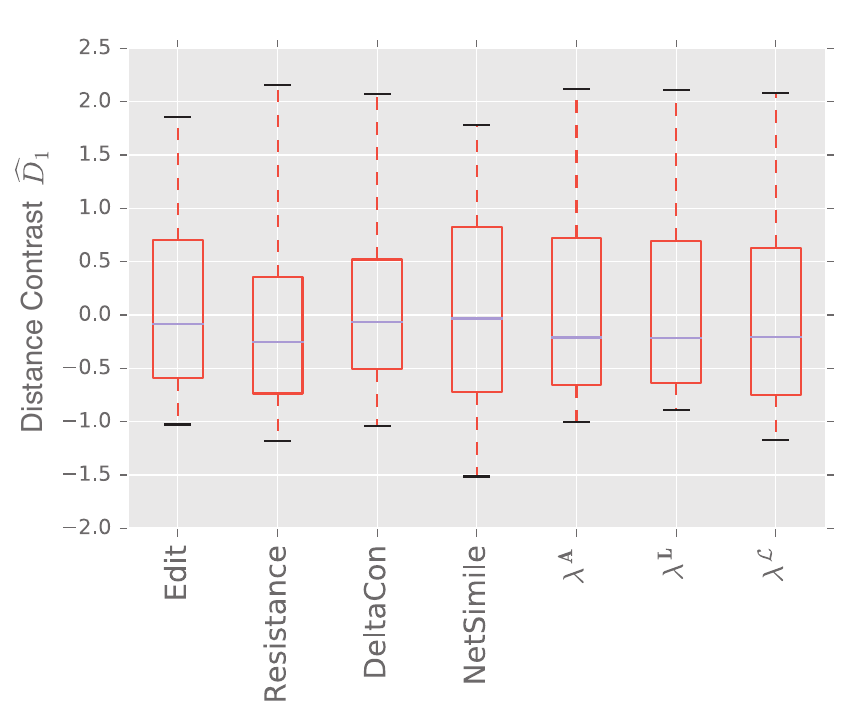}
  }
  \caption{ABIDE data set: distance contrast $\hD_1$ between the weighted
    ASD and control connectomes for a threshold $T=0.5$
    \label{fig:ABIDE3}}
\end{figure}

\begin{figure}[H]
  \centerline{
    \includegraphics[width=0.7\textwidth]{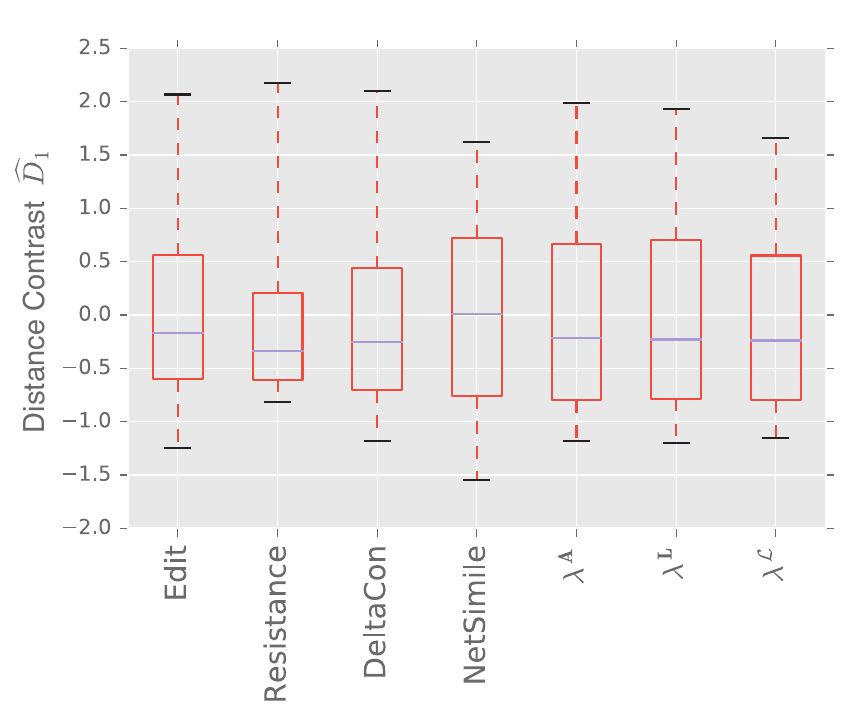}
  }
  \caption{ABIDE data set: distance contrast $\hD_1$ between the weighted
    ASD and control connectomes for a threshold $T=0.8$
    \label{fig:ABIDE4}}
\end{figure}

Fig~\ref{fig:ABIDE1} and Fig~\ref{fig:ABIDE2} display the distance contrast between the set of
unweighted ASD connectomes and the set of control connectomes for two values of the threshold: 0.5
and 0.8 respectively.  Fig~\ref{fig:ABIDE3} and Fig~\ref{fig:ABIDE4} display the distance contrast
between the weighted ASD and control connectomes for two values of the threshold: 0.5 and 0.8
respectively.


Unfortunately, no distances can effectively separate the two ensembles of connectomes. Indeed, the
negative median of the distance contrast $\hD_1$ indicates that the distance between two
connectomes in the ASD and control populations respectively, $D_1$, is on average \textit{lower} than
the average distances between two connectomes from the control population, $\mu_0$ (see \eqref{eq:4}).

This result indicates that the variability in the control population is greater than the contrast
between the two populations. A refined analysis, provided in Section \ref{discussion_fmri}, shows
that the structural differences between the two graph ensembles are localized within subsets of
edges, and cannot be detected when one compares both complete sets of edges. Furthermore, the local
changes in connectivity are of the same order of magnitude as the random local variations present in
these connectomes. For these two reasons, a global comparison using graph metrics seems ineffective
for this problem.

\section*{Discussion
\label{sec:RG_discussion}}
\setcounter{section}{4}\setcounter{subsection}{0} \setcounter{subsubsection}{0}
This section provides an analysis of the numerical simulations and the results of experiments
conducted on real world graphs. Because the numerical simulations were performed using random graph
ensembles, we provide some theoretical justification to explain our findings. The performance of the
distances is studied using a ``multiscale lens'': we organize distances according to the scale at
which they aptly detect changes within a graph. We consider three classes of scales: (1) the fine scale of
the local connectivity, formed by the ego-net; (3) the very large scale associated with communities;
and finally (2) a mesoscale that bridges the scales from the local to the global
scales. Interestingly, this multiscale paradigm has inspired methods to synthesize networks with
guaranteed structural properties at multiple different scales \cite{gutfraind2015}.

For all the graph ensembles studied in this work, we expect random fine scale changes triggered by
the stochastic nature of the models. At the other end of the scale, we expect that certain changes
in connectivity may have dramatic large scale changes.
\begin{figure}[H]
  \centering
  \includegraphics[width=0.6\textwidth]{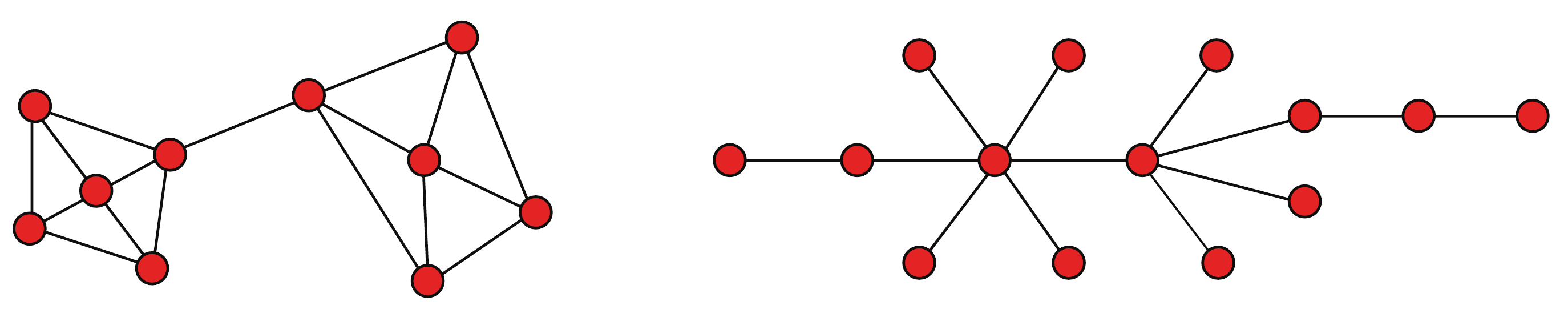}
  \caption{Two significant global structures observed in our experiments. On the
    left is the community structure typical of the stochastic blockmodel. On the
    right is the heavy-tailed degree distribution typical of the preferential
    attachment model.}
  \label{fig:structures}
\end{figure}

Fig~\ref{fig:structures} provides a cartoon
of this concept: changes in edge connectivity of a stochastic blockmodel (left) result in the
creation of a preferential attachment model with the prototypical presence of hubs with very high
degree. 

Finally, the power-law degree distribution of the preferential attachment model suggests
that the graph connectivity involves multiple scales spanning from the finest scale up to the
coarsest scale. Distances adapted to these ``mesoscales'' should be optimal to detect these graphs.
Similarly, we expect that the small world model require the analysis of connectivity at the
mesoscale.
\subsection{The Multiscale detection of random graph ensembles
\label{discussion_ensembles}}
\subsubsection{Detecting large scale changes
\label{sec:large_scales}}
In this study, we focus on experiments where the coarse-scale structure involves the presence (or
absence) of communities. The prototypical ensemble to study the ability of distances to detect
communities is the stochastic blockmodel.

We study a stochastic blockmodel with two partitions of equal size, and we thus expect the second
eigenvalue $\lambda_2$ (of either one of the three matrices) to be the primary distinguishing
spectral feature of the graph. Fig~\ref{fig:SBM_spec} confirms our analysis. We conjecture that $k$
eigenvalues be needed to detect $k$ communities. Indeed, the authors in \cite{Zhang2014} have shown
that the spectrum of the adjacency matrix is composed of two distinct components.  A continuous
spectrum (the bulk) that is centered around 0 is a modified version of the classic semicircle
law. The discrete spectrum is the second component; it is composed of discrete eigenvalues,
distributed away from the continuous spectrum. The number of discrete eigenvalues is equal in number
to the number of communities in the network. The separation between the continuous and the discrete
spectra is what allows our spectral distances to function effectively in detecting community
structure.

\begin{figure}[H]
  \centerline{
    \includegraphics[width=0.7\textwidth]{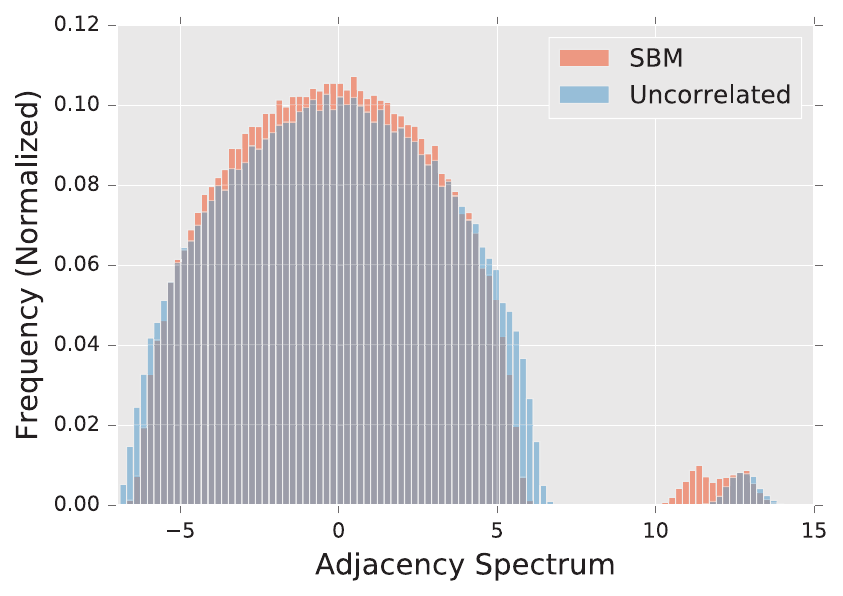}
  }
  \caption{Empirical spectral densities $\lamA$  of the adjacency matrix for the stochastic
    blockmodel (blue) and the uncorrelated random graph (orange). Densities are built from an
    ensemble of 1,000 graphs generated using parameters described in section~\ref{sec:SBM_results}.  
    \label{fig:sbm_spectrum}}
\end{figure}

Fig~\ref{fig:sbm_spectrum} shows the empirical spectral densities of the adjacency matrices
($\lamA$) for the stochastic blockmodel (blue) and the uncorrelated random graph (orange). The
density are well separated around the second largest eigenvalue $\lamA_2$. The bulk of the spectra
for both models overlap significantly, and provide no hope of separating the models from these
eigenvalues. Consequently,  using additional eigenvalues decreases the contrast by including noise in the
comparison (see Fig~\ref{fig:SBM_spec}). 
The use of the spectrum for community partitioning in graphs has a long history (e.g.,
\cite{Luxburg2007} and references therein). Recently, Lee et al. \cite{Lee2014} have proven a
performance bound on the effectiveness of using the first $k$ eigenvectors to partition the graph
into $k$ clusters. In practice, if the graph includes more than two communities of different sizes,
the optimal contrast will require more than the first non trivial eigenvalue.

In summary, we find that when examining global structure, the adjacency spectral distance and \DC
distance both provide good performance. When examining community structure in particular, one need
not employ the full spectrum when using a spectral distance. 
\subsubsection{Detecting mesoscale changes}
In this paper we studied two random graph ensembles whose connectivity structures span several
scales: (1) the preferential attachment model with a non-negligible number of highly connected
vertices (hubs) and a large number of vertices with low degree; (2) the Watts-Strogatz model where
high-degree vertices are extremely unlikely, and where generative rewiring mechanism does not result
in the presence of communities in the graph.

To differentiate graphs based on mesoscale connectivity structures, one should use a spectral
distance computed from either the combinatorial graph Laplacian or the adjacency matrix.\\
~\\

\paragraph{The Combinatorial Graph Laplacian Spectral Distance.} We find that the best
tool for detecting graphs whose degree distribution exhibits polynomial decay \cite{Barabasi1999} is
the combinatorial Laplacian spectral distance. The presence of the degree matrix $D$ in the
Laplacian $\bL=\bD-\bA$ means that comparison of Laplacians is very effective for discerning between
models with radically different degree distributions. Since significant differences between the
degree distributions of the preferential and attachment graphs occur in the tail (i.e. high-degree
vertices), the inclusion of the final few eigenvalues is essential if one wishes to use the
Laplacian spectrum to perform this comparison.

\begin{figure}[H]
  \centerline{
    \includegraphics[width=0.7\textwidth]{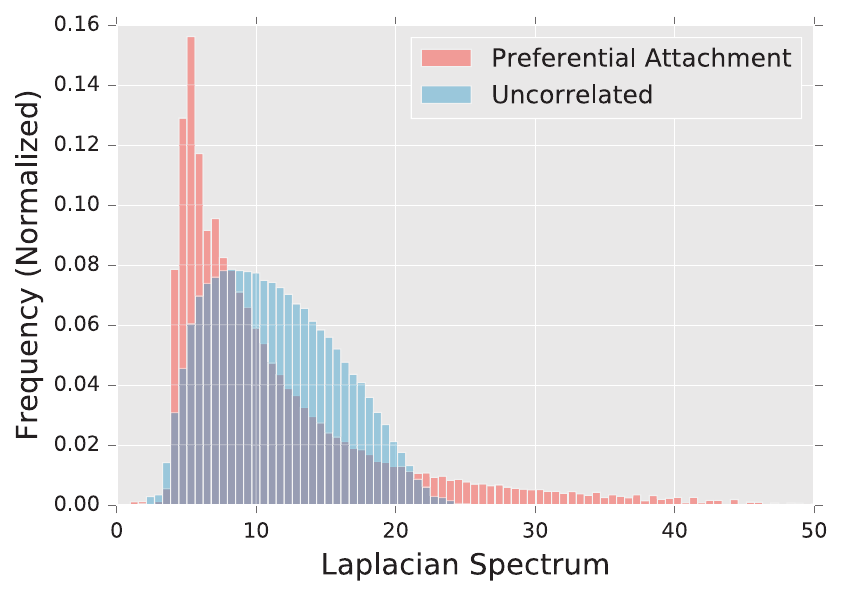}
  }
  \caption{Empirical spectral densities $\lamL$ of the combinatorial Laplacian for the preferential
attachment model (blue) and the uncorrelated random graph (orange). Densities are built from an
    ensemble of 1,000 graphs  generated using parameters described in section~\ref{sec:PA_results}.
    \label{fig:pa_spectrum}}
\end{figure}

Fig \ref{fig:pa_spectrum} displays the empirical spectral densities of the normalized Laplacian
($\lamL$) for the preferential attachment model (blue) and the uncorrelated random graph (orange).

We observe qualitatively, as demonstrated in \cite{Farkas2001}, that the tails of the Laplacian
spectrum of a preferential attachment graph exhibits polynomial decay similar to the tail of the
degree distribution. This is a prime example of the way in which the spectrum of the Laplacian can
be heavily influenced by the degree distribution.

\paragraph{The Adjacency Spectral Distance.} Our findings indicate that the adjacency spectral
distance is the optimal distance for detecting graphs with short average distances, such as the
Watts-Strogatz. Farkas et al. \cite{Farkas2001} argue that the presence of a high number of
triangles is the distinguishing feature of a Watts-Strogatz model. The third moment of the spectral
density of $\bA$ yields the expected number of triangles in a graph, and so one would expect
inclusion of the full spectrum important in detecting the topological signature of this model.

\begin{figure}[H]
  \centerline{
    \includegraphics[width=0.7\textwidth]{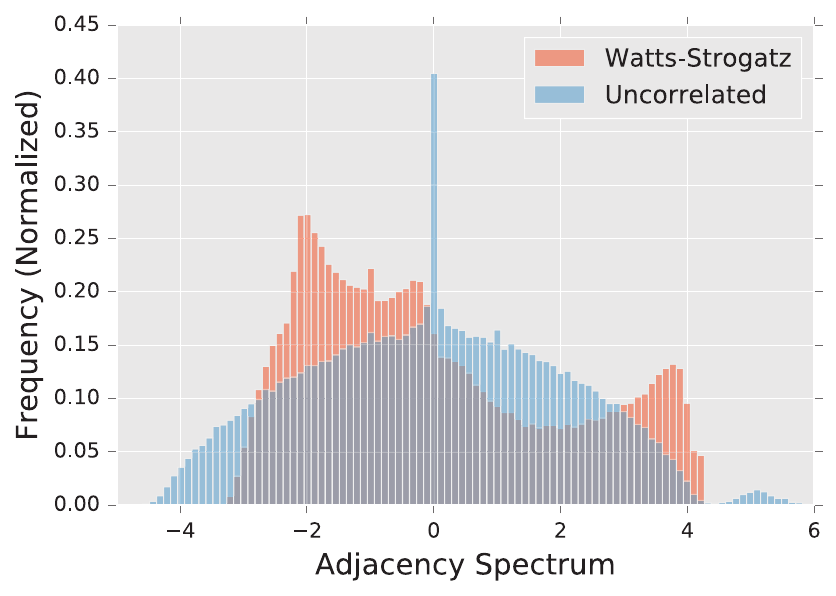}
  }
  \caption{Empirical spectral densities $\lamA$  of the adjacency matrix for the Watts-Strogatz
    model (blue) and the uncorrelated random graph (orange). Densities are built from an 
    ensemble of 1,000 graphs  generated using parameters described in section~\ref{sec:WS_results}.
    The uncorrelated random graph model has a value of $p$ that is smaller than those used in the
    previous models, creating a sharp peak at $\lamA=0$.
    \label{fig:ws_spectrum}}
\end{figure}

This theoretical analysis is confirmed in our experiments. We see in Fig~\ref{fig:WS_k_results} that
inclusion of the large-$k$ (high frequency) eigenvalues is essential to differentiate between the
Watts-Strogatz and the random graph models. Fig \ref{fig:ws_spectrum} confirms that the empirical
spectral density of the Watts-Strogatz model exhibits high skewness, requiring the inclusion of the
bulk of the spectrum to be able to differentiate this model from the random uncorrelated graph (see
Fig~\ref{fig:ws_spectrum}.

A more refined analysis confirms that the very fine scale connectivity, such as the degree
distribution, of the Watts-Strogatz is similar to that of the random graph model, and therefore the
inclusion of the high modes (high $\lamA_k$) decreases the contrast between the two models (see Fig
\ref{fig:WS_k_results}).
\subsubsection{Impact of local structure}
\paragraph{Is the local structure signal or noise?} In this work we consider that the
local scale is defined by the local connectivity at the level of each vertex. Because our study
relies on random graphs ensembles, the local connectivity is intrinsically random. In some of the
graph models, the generative model that leads to the realization of the random graphs induce some
coupling across the scales. The fine scale statistics, such as the degree distribution, become a
``window'' on larger scale patterns of connectivity that happen at multiple scales.

We provide two examples of such phenomena. We first revisit the preferential attachment model (see
Section~\ref{sec:PA_results}). Fig \ref{fig:PA_k_results} shows that the spectral distance based on
the combinatorial graph Laplacian needs the very fine scale, or high frequencies, provided by the
eigenvalues $\lamL_k$ for large $k$ to detect the preferential attachment model. This is
interesting, since this local scale is determined by the degree distribution. Should the null model
mimic the degree distribution of the preferential attachment model (see Section
\ref{sec:DD_PA_results}) then the two graphs become indistinguishable (see Fig
\ref{fig:DD_PA_results}). In this example, the fine scale clearly provides a ``signature'' of the
graph connectivity.

The lattice graph is another extreme example of where the local connectivity structure can be used
to identify the graph. The lattice graph includes cycles of any size (starting with length 4). As a
result, the spectral distances all benefit from increasing the number of eigenvalues used to compare
the graphs (see Fig \ref{fig:lattice_k_results}).
\vspace*{1em}\\

On the other hand, random fluctuations can also be a source of uninformative noise when comparing
graphs. The results of Section \ref{sec:SBM_results} illustrate this fact. The Laplacian spectrum is
unable to distinguish between the stochastic blockmodel and the uncorrelated random graph, while the
normalized Laplacian distinguishes them well. The difference between these two matrix
representations is that normalization removes degree information, which is not informative in this
particular model (see Fig \ref{fig:SBM_spec}).

A similar problem arise when we apply the resistance distance to the stochastic blockmodel; as
discussed in the previous section, the resistance distance is disproportionately influenced by local
structure, and is unable to discern the global structure of the graph over local
fluctuations. Interestingly, \DC does not appear to suffer from local fluctuations as much as the
resistance distance. This could be due to the structure of the matrix $\mathbf{S}$ that \DC uses to
represent the graph, or due to the use of the Matusita distance rather than the $\ell_1$ or $\ell_2$
norm to compare the resulting matrices (for more discussion of this, see Sections 2.2 and 3.1 in
\cite{Koutra2014}).

In summary, it is essential to determine whether local topological features are of interest in the
comparison problem at hand; inclusion of locally targeted distance measures can hinder the
performance of graph distances in cases where local structure is noisy and uninformative. However,
if local structure is ignored, one can often omit essential structural information about the graphs
under comparison.
\subsection{Real world networks 
  \label{discussion_real_networks}}
Experiments performed on random graph ensembles provide a mechanism to gauge the ability of each
distance to detect changes in structural features that are prototypical of the corresponding
ensembles (e.g., communities, clustering coefficients, power law degree distribution, etc.).
Specifically, this analysis lends itself to a systematic exploration of an experimental version of
the  two-sample test problem where we compare two populations of random graphs using a distance
statistic, and we experimentally test whether both populations could be generated by the same probability
distribution.

In this context, we explore the two-sample test problem in neuroscience, and compared two
populations of functional brain networks. Signal-to-noise is a ubiquitous problem in analyzing
actual graph data, and is particularly notable in building connectivity networks of human brain
activity (see e.g., \cite{Burgess2016}). Accordingly, the results of our data experiments show that
in the presence of real-world noise levels, many of these distances fail to distinguish subtle
structural differences. In the face of this, we examine more targeted analysis techniques that may
be applied in such a situation.

A related question concerns the change point detection scenario for a dynamic graph, where we detect
significant changes between adjacent time steps using a distance \cite{Holme2012}. The first
experiment suggests that the tools that perform the most consistently in the two-sample test problem
(the spectral distances) are unreliable in the change detection experiment. This experiment is
interesting because it allows us to evaluate distances in a context where graphs exhibit significant
volume fluctuations, a situation that we did not encounter in our numerical studies.
\subsubsection{Primary school face to face contact
\label{edge_perturb}}
The primary school face to face contact dataset (\ref{sec:primary_xp}) provides a real-world
example to evaluate the performance of distances in the context of a dynamic network.

The purpose of the analysis is to assess whether distances can detect changes in the topology
coupled with the hidden events that control the network topology and connectivity (such as those that
occur during the lunch period). We are also interested to verify if distances are robust against
random changes within each classroom that do not affect the communication between the classes (e.g.,
see Fig.~\ref{fig:contact_graphs} at times 10:50 a.m., 10:57 a.m.). 

The most remarkable conclusion of this particular experiment is that although the spectral distances
are very efficient and stable for the purposes of comparing two random graphs sampled from distinct
probability models (see section \ref{sec:results_graph_models}), these distances perform poorly in
the context of change point detection (see Fig \ref{fig:primary_school3}). In contrast, the
resistance distance can detect subtle topological changes that are coupled to latent events that
dynamically modify the networks. The resistance distance remains impervious to random local changes,
which do not affect the large scale connectivity structure (see Fig \ref{fig:primary_school1}).

Unlike the analysis of random graphs, where the volume of the two graphs were always the same, the
volume of the dynamic network changes rapidly, and therefore the edit distance exhibits significant
changes throughout the school day. While the edit distance can reliably monitor large scale changes
in the graph volume, it entirely misses the significant events that disrupt the graph topology:
onset and end of morning recess, onset of first lunch, end of second lunch (see Fig
\ref{fig:primary_school2}).

With the help of Fig.~\ref{fig:contact_graphs} (the snapshots are obtained from the movie available
on \cite{sociopattern}), we analyze some of the most significant differences between the three
distances.

At time 10:20 a.m., $\hD_R$ changes abruptly (see Fig~\ref{fig:primary_school1}) as a result
of a massive increase in the number of contacts between students in the second, third, and fourth
grades (see Fig~\ref{fig:contact_graphs}). While $\hD_E$ and $\hD_{DC}$ also register
this change, they are less sensitive to the merging of the communities than the
resistance-perturbation distance.

The significant difference between $\hD_R$ and the two other distances before 11:00 a.m.  (see
Fig~\ref{fig:primary_school1}) is also very interesting. Because the recess period is winding down
from 10:50 a.m. to 10:57 a.m., the number of contacts within each class decreases very significantly
(especially in the two second grade classes, see Fig~\ref{fig:contact_graphs}). These {\em within
  the classes} changes are easily detected by $\hD_{DC}$ and $\hD_E$, which continue to grow during
this time interval. However, there are only very few changes in the contacts {\em between the
  classes} during that period (see Fig~\ref{fig:contact_graphs}). Consequently, $\hD_R$ becomes very
small (see the significant dip of $\hD_R$ shortly before 11:00 a.m. in
Fig~\ref{fig:primary_school1}).

The resistance-perturbation distance $\hD_R$ is also able to detect the dissolution of the
classes at 11:57 a.m.  just before the official lunch period, as the students are running outside of
the classrooms into the hallway. The random appearance of the connectivity (see
Fig~\ref{fig:contact_graphs}) reflects the activity of the students. This is significant, because
this event happens before the number of contacts increases (the edit distance jumps right after
12:p.m., see Fig~(see Fig \ref{fig:primary_school2})

The geometry of the graph at 12:13 p.m. (see Fig~\ref{fig:contact_graphs}) is indicative of the fact
that half of the students take their lunch in the cafeteria, while the other half play in the
courtyard. In spite of the fact that the DeltaCon distance is still large, the contact network is in
fact in a large scale stable topological configuration, leading to a small $\hD_R$.

In fact, $\hD_R$ can also detect the pattern of activity associated with the second lunch
period at 12:54 p.m. (see Fig~\ref{fig:primary_school1}). Because this is only a reconfiguration of the
network, the edit distance is oblivious to these changes (see Fig~\ref{fig:primary_school2}).

Finally, we note that the resistance distance can detect the early regrouping of the students around
1:46 p.m. (see Fig~\ref{fig:primary_school1}), according to their classroom (see
Fig~\ref{fig:contact_graphs}), before the end of the lunch period.

Interestingly, $\hD_R$ can also detect the small number of students who are late going back
to their class between 2:00 p.m. and 2:03 p.m. (see Fig~\ref{fig:contact_graphs}). The edit and
DeltaCon distances are not affected by the removal of these cross-community edges (see
Figs~\ref{fig:primary_school2},\ref{fig:primary_school1} respectively), but the resistance distance
is greatly influenced by these changes, and thus $\hD_R$ is very large (see
Fig~\ref{fig:primary_school1}).\\~\\

These structural changes are of a global nature. In Sections \ref{sec:SBM_results} and
\ref{sec:large_scales} we saw that the spectral distances were more effective than the matrix
distances to detect large-scale differences between the following two graph ensembles: the stochastic
blockmodel and the uncorrelated random graph model. Because the dynamic network is comprised of
communities, a naive analysis would suggest that spectral distances should also outperform the
resistance perturbation distance and \textsc{DeltaCon}.

A refined analysis demonstrates that if a dynamic network is composed of communities, the resistance
perturbation provides the ideal solution to the change point detection problem, by effectively
ignoring the rapid random fluctuations at the node level, while remaining sensitive to changes in
connectivity between communities.

This analysis requires that we review the expression of the resistance-perturbation in terms of the
eigenvalues and eigenvectors of the normalized graph Laplacian \cite{Monnig2016}.  The present
authors derived in \cite{Monnig2016}, a closed-form expression of the resistance-perturbation
distance between a graph and a rank-one perturbation of that graph, wherein a single edge has
changed. This theoretical analysis is useful indeed, because it provides the baseline scenario to
compare various changes in connectivity in the context of the change point detection scenario for a
dynamic graph.

We recall that $\bfi_k$ denotes the $k^{\text{th}}$ eigenvector of the normalized graph Laplacian,
with the eigenvalues organized as $0 = \lamNL_1 \le \ldots \le \lamNL_n$.  We also denote by
$\drpo(G,G+\dw{i_0j_0}) $ the resistance perturbation distances between a graph $G$, and the graph
obtained from $G$ by a perturbation $\dw{i_0j_0}$ to the edge $(i_0,j_0)$, $G+\dw{i_0j_0}$.
\begin{theorem}[resistance-perturbation after edge modification \cite{Monnig2016}]
  If $G+\dw{i_0j_0}$ is the graph obtained from $G$ by a perturbation $\dw{i_0j_0}$ to
  the edge $(i_0,j_0)$, then 
  \begin{equation}
      \drpo(G,G+\dw{i_0j_0}) = \frac{2 n  \left\lvert \dw{i_0j_0} \right\rvert}    {1 + \dw{i_0j_0}
        R_{i_0j_0}}\;  \sum_{k=2}^n \frac{1}{(\lamNL_k)^2} \left[\bfi_k(i_0) - \bfi_k(j_0) \right]^2.
    \label{edge_mod_eq}
  \end{equation}
  \label{edge_mod_thm}
\end{theorem}
As expected, the  term
\begin{equation}
  \frac{\dw{i_0j_0} }{1 + \dw{i_0j_0} R_{i_0j_0}},
  \label{ratio_drpo}
\end{equation}
controls the size of the resistance-perturbation distance $\drpo(G,G+\dw{i_0j_0})$.

More interestingly, the sum $\sum_{k=2}^n \left[\bfi_k(i_0) - \bfi_k(j_0) \right]^2 /(\lamNL_k)^2$
in (\ref{edge_mod_eq}) provides the ``frequency response'' of the graph to the perturbation. This
response can be analyzed as follows.  For small $k$, the eigenvalues $\lamNL_k$ are small, and the
corresponding eigenvectors $\bfi_k$ ``oscillate'' very slowly on the graph, i.e.
$\bfi_k(i_0) - \bfi_k(j_0) \approx 0$ unless $i_0$ and $j_0$ belong to different nodal regions. In
this latter case, the effect of the edge perturbation $\dw{i_0j_0}$ will be maximal. An example of
this phenomenon corresponds to the primary school face-to-face contact network, where each classroom
forms a densely connected community. The classrooms are weakly connected to one another. For the
same $\dw{i_0j_0}$, $\drpo(G,G+\dw{i_0j_0})$ will be maximal if $i_0$ and $j_0$ are in different
classrooms, and will be very small if the two nodes belong to the same classroom.

For large $k$, eigenvectors $\bfi_k$ ``oscillate'' very quickly on the graph, making it
difficult to estimate the contribution of $\left[\bfi_k(i_0) - \bfi_k(j_0) \right]^2$. This issue is
mitigated by the fact that the weights $1/(\lamNL_k)^2$ are relatively small, since the eigenvalues
$\lamNL_k$ are large.\\~\\

In summary, the resistance-perturbation provides a multiscale analysis that automatically
de-emphasize the random variability at the very fine scales, to wit the distance ``denoises'' the
graph dynamic (see also \cite{meyer2014,cheung2018} for similar analyses).

We note the existence of topological distance \cite{saggar2018,geniesse2019,moo2019} that also provide ``multiscale
distances'' through a filtration process. These topological distances go beyond the scope of the current study that
focuses on geometric distances.
\color{violet}
\subsubsection{\color{violet}
European Union  Emails}
The analysis of the pattern of connectivity of email exchanges between the members of a European research institute
provides a second example of dynamic network. The first observation is that the rate of changes in the number of emails
(volume of each graphs) appears to be influenced by the calendar of the European Parliament (see
Fig. \ref{fig:eu-emails-vol}). As suspected by the authors in \cite{hajij18}, the activity in this dataset appears to be
influenced by a series of events at the European Parliament in Brussels and Strasbourg. This conjecture is confirmed by
the analysis of the spectral distances (see Fig. \ref{fig:eu-emails-spectral}): the three distances show large values
whenever the session of the Parliament is resumed (e.g., 3 December, 2003; 19 April 2004, 13 October 2004, etc.) or
adjourned (e.g., 18 December 2003, 1 April 2004, 28 October 2004, etc.) More interestingly, all the distances become
suddenly very large during the entry into office of the new 2004-2009 European Commission (22-26 November, 2004).

Unlike all the spectral distances, the resistance distance and the \DC distance are able to detect the election of Jose
Barroso as President of the European Commission (22 July 2004) (see Fig. \ref{fig:eu-emails-mat}). We had already
observed and analysed the performance of these two distances in the context of the face-to-face contact networks. This
experiment confirms that the resistance perturbation provides a very sensitive statistic to detect changes in dynamic
networks.

\subsubsection{Functional brain connectivity
\label{discussion_fmri}}
The experiments in section \ref{sec:fmri_results} leave the choice of the global distance open. In
this section, we gain further insight into the analysis of this dataset, and demonstrate that local
changes in the connectivity of functional brain networks can indeed be detected. These minute
changes require assigning a more robust weight along the edges, and designing a distance between
graphs that can be tuned to respond to local changes at specific scales (a weighted spectral
distance). The implementation of these ideas is beyond the scope of the paper.

\begin{figure}[H]
  \centering
  \includegraphics[width=0.7\textwidth]{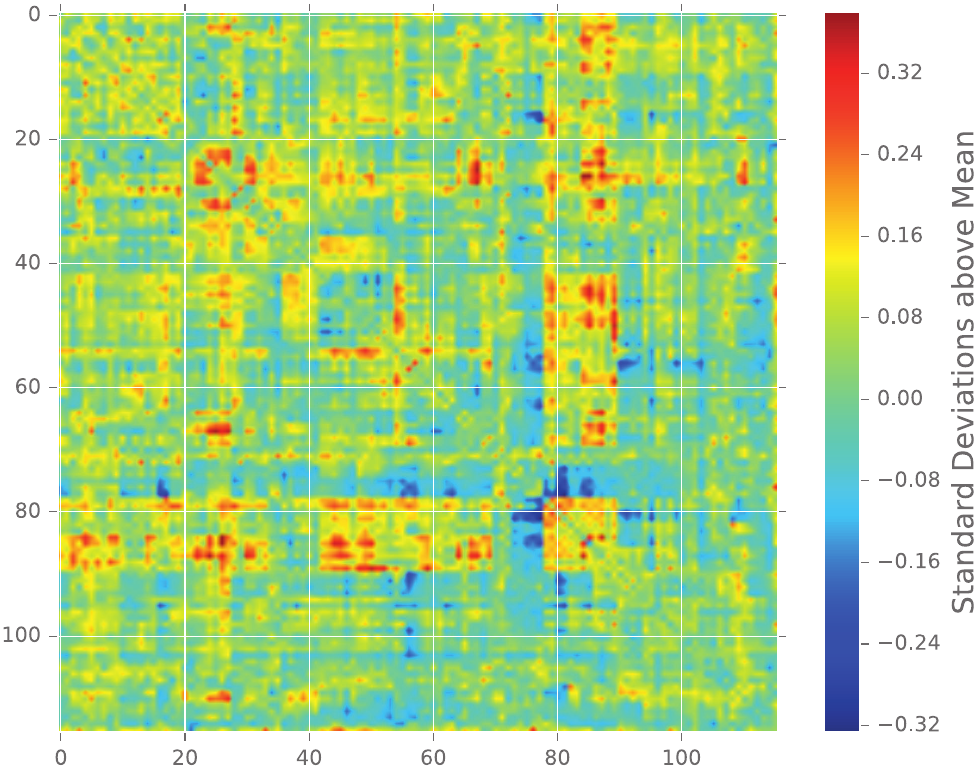}\\
  \caption{Contrast $\hD_{i,j}$ for each pair of regions $(i,j)$ in the atlas AAL computed between the ASD
    subjects and the controls (see \eqref{eq:contrast_fmri}).}
  \label{fig:asd_corr}
\end{figure}

In order to gain some understanding into the inability of the graph distance to differentiate
between the ASD patients and the controls, we revisit the original data, and compute the following
contrast for each pair of nodes $(i,j)$ in the network
\begin{equation}
  \hD_{i,j} \eqdef \frac{\wE{\rho^{1}_{i,j}} -
    \wE{\rho^{0}_{i,j}}}{\sigma\left(\rho^{0}_{i,j}\right)},
  \label{eq:contrast_fmri}
\end{equation}
where $\wE{\rho^{1}_{i,j}}$ is the sample mean correlation between regions $i$ and region $j$ of the
brain atlas, computed over all ASD subjects (population 1). Similarly, $\wE{\rho^{0}_{i,j}}$ and
$\sigma\left(\rho^{0}_{i,j}\right)$ are the sample mean and sampled variance, respectively, of the
correlations between regions $i$ and region $j$ of the brain atlas, computed over all controls
(population 0). Fig \ref{fig:asd_corr} displays the contrast $\hD_{i,j}$ for all pair of regions in
the AAL atlas, $i,j=1,\ldots 116$. 

To ease the interpretation of this matrix, consider for instance a value of $\hD_{i,j} =0.25$. This
value indicates that (on average) the functional correlation $\wE{\rho_{i,j}^{1}}$ between regions
$i$ and $j$ of ASD subjects is above the (average) correlation of the controls,
$\wE{\rho_{i,j}^{1}}$, by $0.25 \times$ the standard deviation of the correlation (of the controls),
$\sigma\left(\rho^{0}_{i,j}\right)$.

A visual inspection of Fig \ref{fig:asd_corr} highlights the presence of localized high contrast
between the two populations. A closer examination reveals that ASD subjects are 
underconnected in regions 73 through 77, and are overconnected in regions 79 and 84 through
89. Table \ref{tab:AAL_regions} provides a list of regions that are show anomalous functional
connectivity.
\begin{table}[H]
  \centering
  \begin{tabular}{|l|l|l|}
    \hline
    \textbf{Label} & \textbf{Region} & \textbf{Connection}\\
    \hline
    73 & L. Putamen & Underconnected\\
    74 & R. Putamen & Underconnected\\
    75 & L. Globus Pallidus & Underconnected\\
    76 & R. Globus Pallidus & Underconnected\\
    77 & L. Thalamus & Underconnected\\
    79 & R. Transverse Temporal Gyrus & Overconnected\\
    84 & R. Superior Temporal Lobe & Overconnected\\
    85 & L. Middle Temporal Gyrus & Overconnected\\
    86 & R. Middle Temporal Gyrus & Overconnected\\
    87 & L. Middle Temporal Pole & Overconnected\\
    88 & R. Middle Temporal Pole & Overconnected\\
    89 & L. Inferior Temporal Gyrus & Overconnected\\
    \hline
  \end{tabular}
  \caption{Regions with anomalous connectivity. Correspondence between labels and regions
    is established via the Automated Anatomical Labelling atlas \cite{Tzourio2002}.}
  \label{tab:AAL_regions}
\end{table}
Fig ~\ref{fig:asd_corr} confirms that there exist structural differences between the connectomes of
ASD subjects and controls. Unfortunately, our analysis shows that these differences are smaller than
one standard deviation of the correlation of the controls $\sigma\left(\rho^{0}_{i,j}\right)$ (see
the color bar in Fig \ref{fig:asd_corr}). To further aggravate this situation, we note that the
pattern of anomalous connectivity are isolated, while the vast majority of correlations are very
close to zero (green cells in Fig~\ref{fig:asd_corr}). We conclude that the low amplitude of the
contrast $\hD_{i,j}$ and its sparsity contribute to our inability to use graph distances to detect
significant changes between the connectomes of the two populations (see also \cite{rasero2018} for a
detailed analysis of regional connectivity). We note that others have reported similar findings
\cite{Redcay2013,Hull2016}. 
\section*{Conclusion
  \label{sec:conclusion}} \setcounter{section}{5}\setcounter{subsection}{0}
\setcounter{subsubsection}{0}
The success of statistical machine learning relies on the construction of sophisticated spaces of
signals (functional spaces) wherein properties of algorithms can be rigorously evaluated. The core
of the analysis usually relies on the existence of bases that reveal the properties of the class of
functions of interest. There currently is no equivalent for the study of graph ensembles. 

In this paper, we considered existing ensembles of random graphs as prototypical examples of certain
graph \emph{structures}, which are the building blocks of existing real world networks. These
ensembles were used to rigorously analyze various graph distances in the context of the
two-sample test problem.

Specifically, we studied the ability of various distances to compare two samples randomly drawn from
distinct ensembles of graphs. We investigated the relationship between the families of graph
ensembles, the structural features characteristic of these ensembles, and the sensitivity of the
distances to these characteristic structural features. The performance of the distances is studied
using a ``multiscale lens'': we organize distances according to the scale at which they aptly detect
changes within a graph. We consider three classes of scales: (1) the fine scale of the local
connectivity, formed by the ego-net; (3) the very large scale associated with communities; and
finally (2) a mesoscale that bridges the scales from the local to the global scales.

We concluded our study with experiments conducted on real-world networks, where we study the
two-sample test problem for networks of functional brain connectivity, and we detected change points
in a dynamic network of face-to-face contacts.
\subsection{Recommendations}
Throughout this study, we observed that the adjacency spectral distance (see Sections
\ref{sec:results_graph_models} and \ref{sec:results_real_networks}) exhibits good performance across
a variety of scenarios, making it a reliable choice for a wide range of problems. Spectral distances
also exhibit practical advantages over matrix distances, as they can inherently compare graphs of
different sizes and can compare graphs without known vertex correspondence. The adjacency spectrum
in particular is well-understood, and is perhaps the most frequently studied graph spectrum; see
e.g., \cite{Farkas2001, Flaxman2003}. Finally, fast, stable eigensolvers for symmetric matrices are
ubiquitous in modern computing packages such as ARPACK, NumPy, and Matlab, allowing for rapid
deployment of models based on spectral graph comparison.\footnote{The Python library
  \texttt{NetComp} \cite{netcomp19} further simplifies the application of these tools to practical problems; see the
  appendix for more details.} Furthermore, randomized algorithms for matrix decomposition allow for
highly parallelizable calculation of the spectra of large graphs \cite{Halko2011}.

However, the utility of the adjacency spectral distance is not general enough to simply apply it to
any given two-sample test or anomaly detection problem in a naive manner. A prudent practitioner
would combine exploratory structural analysis of the graphs in question with an ensemble approach in
which multiple distance measures are considered simultaneously, and the resulting information is
combined to form a consensus. Such systems are commonplace in problems of classification in machine
learning, where they are sometimes known as ``voting classifiers'' (see e.g., \cite{Roli2001}).

In this study, we have been comparing graphs of equal volume (in expectation). In
situations where the graph volume varies drastically (e.g., see Section \ref{sec:primary-school}),
the process of choosing a graph comparison tool may differ significantly. The situation reverses
when we look at the problem of detecting change points in a dynamic graph (see Section
\ref{sec:primary-school}). In this scenario, the matrix distances proved most effective in detecting
changes in the latent variables controlling the network dynamics. The spectral distances, on the
other hand, were so noisy as to be useless.  When trying to detect change points in a dynamic graph,
one computes the distance between consecutive time steps. In this scenario the two graphs being
compared share many more edges than in the two-sample test. As demonstrated in section
\ref{edge_perturb}, the resistance perturbation distance, or \DC, yield exemplary performance. We
note that we found that raw fluctuations in graph volume did not yield useful information about the
latent processes that triggered changes in graph connectivity.

\begin{figure}
  \centerline{
    \includegraphics[width=0.8\textwidth]{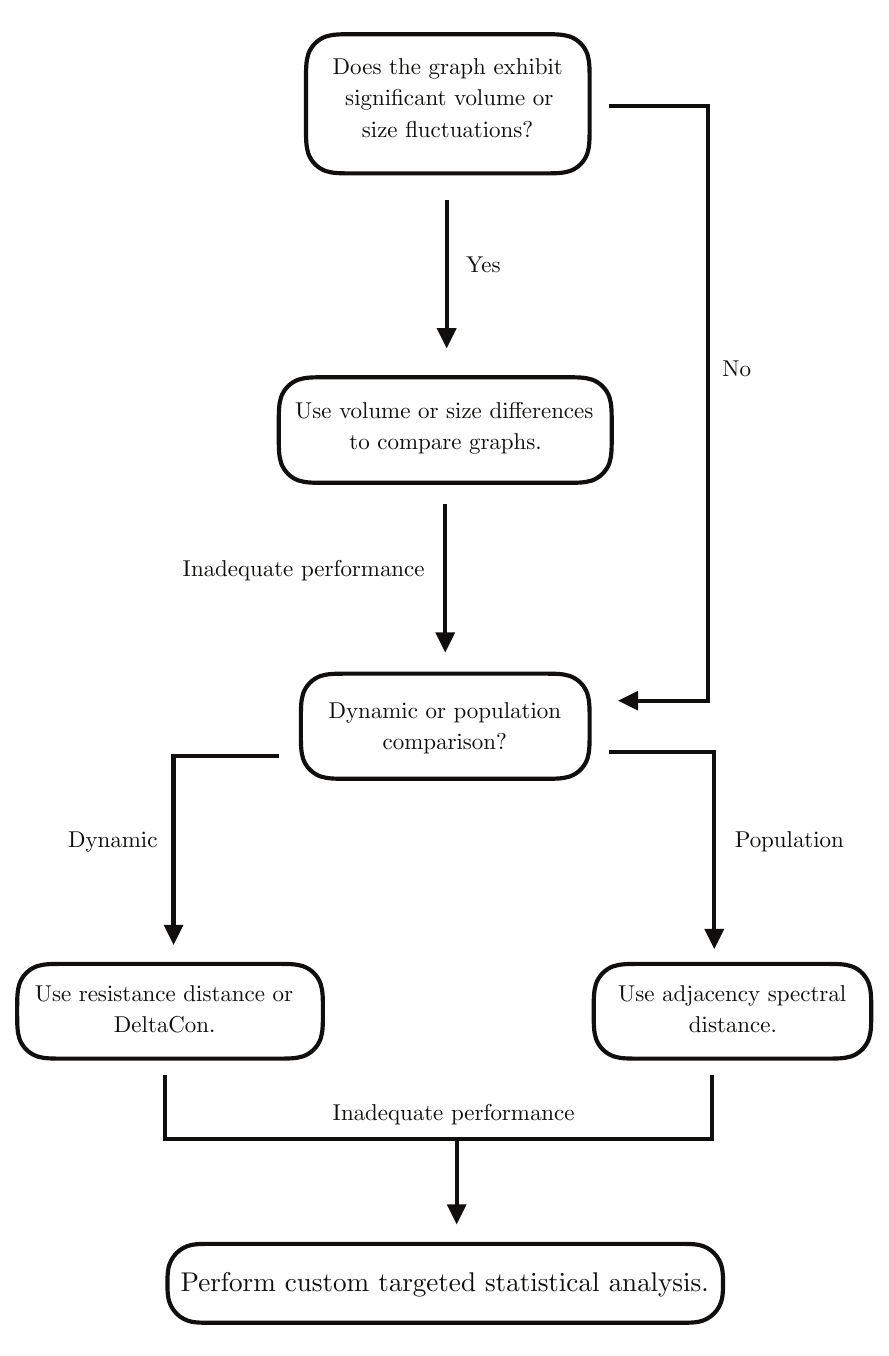}
  }
  \caption{Flow chart summarizing the suggested decision process for applying
    distance measures in empirical data.
    \label{fig:flowchart}}
\end{figure}

Based on the results of our experiments, we provide a suggested decision process in Fig
\ref{fig:flowchart}. If the graphs to be compared exhibit differences in volume or size, then these
should be examined to see if they hold predictive power, as they are simple and efficient to
compute. If they prove ineffective, then one must consider the setting. In a dynamic setting, in
which a dynamic graph is being compared at subsequent time steps, then we recommend using matrix
distances based on the results of Section \ref{sec:primary-school}. If one is comparing graphs to
determine whether a sample belongs to a given population, then the adjacency spectral distance is
the most reliable, as Sections \ref{sec:results_graph_models} and \ref{sec:results_real_networks}
demonstrate. Finally, if none of these approaches give adequate performance, then a more targeted
analysis must be performed, such as the edge-wise statistical comparison of weights in Fig
\ref{fig:asd_corr}. The particular design of this analysis is domain specific and highly dependent
upon the nature of the data.

\appendix
\section*{Notation}

For reference, in Table~\ref{tab:notation} we provide a table of notation used
throughout the paper.

\begin{table}[h]
  \centering
  \begin{tabular}{|l|l|}
    \hline
    $G$ & Graph \\
    $V$ & Vertex set, taken to be $\{1,2,\ldots,n\}$ \\
    $E$ & Edge set, subset of $V \times V$ \\
    $W$ & Weight function, $W:E\rightarrow\mathbb{R}^+$\\
    $n$ & Size of the graph, $n = |V|$\\
    $m$ & Number of edges, $m = |E|$ \\
    $d_i$ & Degree of vertex $i$ \\
    $\bD$ & Degree matrix (diagonal) \\
    $d(\cdot,\cdot)$ & Distance function \\
    $\bA$ & Adjacency matrix \\
    $\bL$ & Laplacian matrix \\
    $\bcL$ & normalized Laplacian matrix (symmetric) \\
    $\lamA_i$ & $i^\text{th}$ eigenvalue of the adjacency matrix\\
    $\lamL_i$ & $i^\text{th}$ eigenvalue of the Laplacian matrix\\
    $\lamNL_i$ & $i^\text{th}$ eigenvalue of the normalized Laplacian matrix\\
    $\cG_{\{0,1\}}$ & The \{null,alternative\} population of graphs\\
    $G_{\{0,1\}}$ & Sample graph from $\cG_{\{0,1\}}$\\
    $\cD_0$ & Distribution of distances between graphs in null population \\
    $D_0$ & Sample from $\cD_0$\\
    $\cD_1$ & Distribution of distances $d(G_0,G_1)$\\
    $D_1$ & Sample from $\cD_1$\\
    $\hat {\cD}_1$ & Distribution of the contrast $\cD_1$ between $D_0$ and $D_1$, (see \eqref{eq:4})\\
    $\hat D_1$ & Sample from $\hat{\cD}_1$ \\
    $G(n,p)$ & Uncorrelated random graph with parameters $n$ and $p$ \\
    $(n,p,q)$ & Parameters for stochastic blockmodel\\
    $(n,l)$ & Parameters for preferential attachment model, with $1<l\leq n$\\
    $(n,k)$ & Parameters for Watts-Strogatz graph, with $k<n$ even\\
    \hline
  \end{tabular}
  \caption{Table of commonly used notation.   \label{tab:notation}}
\end{table}

\section*{NetComp: network comparison in python
\label{sec:netcomp}}
\setcounter{section}{8}
\texttt{NetComp} is a Python library that implements the graph distances
studied in this work. Although many useful tools for network construction and
analysis are available in the well-known \texttt{NetworkX} \cite{Hagberg2008},
advanced algorithms such as spectral comparisons and \DC are not
present. \texttt{NetComp} is designed to bridge this gap. 
\subsection{Design consideration}

The guiding principles behind the library are 

\begin{enumerate}

\item \textbf{Speed}. The library implements algorithms that run in linear or
  near-linear time, and are thus applicable to large graph data
  problems.\footnote{See below regarding the implementation of exact and
    approximate forms of \DC and the resistance distance.}

\item \textbf{Flexibility}. The library uses as its fundamental object the
  adjacency matrix. This matrix can be represented in either a dense
  (\texttt{NumPy} matrix) or sparse (\texttt{SciPy} sparse matrix) format. Using
  such a ubiquitous format as fundamental allows easy input of graph data from a
  wide variety of sources.

\item \textbf{Extensibility}. The library is written so as to be easily extended
  by anyone wishing to do so. The included graph distances will hopefully be
  only the beginning of a full library of efficient modern graph comparison
  tools that will be implemented within \texttt{NetComp}.

\end{enumerate}

\texttt{NetComp} is available via the Python Package Index, that is most
frequently accessed via the command-line tool \texttt{pip}. The user can install
it locally via the shell command
$$\text{\texttt{pip install netcomp}}.$$
As of writing, the library is in alpha. The approximate (near-linear) forms of \DC and the
resistance distance are not yet included in the package. Both algorithms have an quadratic-time
exact form that is implemented. Those interested can download the source code and contribute (by
adding the distance of their choice) at \url{https://www.github.com/peterewills/netcomp}.

\section*{Author contributions}
Writing – Original Draft: P.W.; Writing – Review \& Editing: P.W. and F.G.M.; Conceptualization:
P.W. and F.G.M.; Investigation: P.W. and F.G.M.; Methodology: P.W. and F.G.M; Formal Analysis:
P.W. and F.G.M.; Software: P.W.
\nolinenumbers
\section*{Acknowledgments}
The authors are grateful to the anonymous reviewers and the Academic Editor  for their insightful comments and suggestions that greatly improved
the content and presentation of this manuscript.

\end{document}